\documentclass[useAMS,usenatbib,usegraphicx,referee]{mn2e}
\newcommand{\bugref}{\bibitem[\protect\citeauthoryear{dummy }{1893}]{dum}}

\title[Radio Circular Polarization in Helical B Fields in Eight AGN]
{Radio Circular Polarization Produced in Helical
Magnetic Fields in Eight Active Galactic Nuclei}

\author[D. C. Gabuzda et al.]{D. C. Gabuzda,$^{1}$
V. M. Vitrishchak$^{2}$, M. Mahmud$^{1}$ \& S. P. O'Sullivan$^{1}$ \\
$^{1}$Physics Department, University College Cork, Cork, Ireland \\
$^{2}$Sternberg Astronomical Institute, Moscow State
University, Universitetski{\u\i} prospekt 13, Moscow, 119992
Russia}

\begin{document}

\date{Released 2007 Xxxxx XX}
\pagerange{\pageref{firstpage}--\pageref{lastpage}} \pubyear{2007}
\maketitle
\label{firstpage}
\begin{abstract}
Homan \& Lister (2006) have recently published circular-polarizatio
(CP) detections for 34 objects in the MOJAVE sample -- a set of
bright, compact AGN being monitored by the Very Long Baseline Array at
15~GHz. We report the detection of 15-GHz parsec-scale CP in two
more AGN (3C345 and 2231+114), and confirm the MOJAVE detection of
CP in 1633+382.
It is generally believed that the most likely mechanism for the
generation of this CP is Faraday conversion of linear polarization
to CP. A helical jet magnetic-field ({\bf B}-field) geometry can
facilitate this process --
linearly polarized emission from the far side of the jet is
converted to CP as it passes through the magnetised
plasma at the front side of the jet on its way toward the observer. In
this case,
the sign of the generated CP is essentially determined by the
pitch angle and helicity of the helical {\bf B} field.
We have determined the pitch-angle regimes and helicities of the helical jet
{\bf B} fields in 8 AGN for which parsec-scale CP has been detected,
and used them to predict the expected CP signs for these AGN if the
CP is generated via conversion in these helical fields. We have
obtained the intriguing result that our predictions agree with the 
observed signs in all eight cases, provided that the longitudinal {\bf B}-field
components in the jets correspond to South magnetic poles. This
clearly non-random pattern demonstrates
that the observed CP in AGN is directly associated with the presence of
helical jet {\bf B} fields. These results
suggest that helical {\bf B} fields are ubiquitous in
AGN jets.
\end{abstract}

\section{Introduction}

The radio emission of core-dominated, radio-loud Active Galactic Nuclei
(AGN) is synchrotron radiation
generated in the relativistic jets that emerge from the nucleus of the
galaxy, presumably along the rotational
axis of a central supermassive black hole. Synchrotron radiation can be
highly linearly polarized, to $\simeq 75\%$ in the case of a uniform magnetic
({\bf B}) field (Pacholczyk 1970), and linear polarization (LP) observations
can yield unique information about the orientation and degree of order of the
{\bf B} field in the synchrotron source, as well as the distribution of
thermal electrons and the {\bf B}-field geometry in the immediate vicinity
of the AGN (e.g., via Faraday rotation of the plane of polarization).

After a number of early circular-polarization (CP) 
studies for various samples of AGN in the 1970's and
1980's (Gilbert \& Conway 1970; Conway et al. 1971; Roberts et al. 1975;
Weiler \& Wilson 1977; Weiler \&
de Pater 1983; Komesaroff et al. 1984), little
observational work was done in this area until the late 1990's, when the
Australia Telescope Compact Array was optimised for
CP measurements by Rayner et al. (2000),
and techniques for deriving CP information on parsec
scales were developed by Homan and his
collaborators (Homan \& Wardle 1999; Homan, Attridge \& Wardle 2001)
using the NRAO\footnote{The National Radio Astronomy Observatory of
the USA is operated by Associated Universities, Inc., under
co-operative agreement with the US NSF.} Very Long Baseline Array (VLBA).
The high resolution provided by these latter observations makes it possible
to more precisely localize the regions of CP within the source, and thereby
more accurately estimate the degree of CP in these regions. Certain
tendencies became clear from the very first VLBA CP measurements:
the CP was nearly always coincident with the VLBI core, with the degrees
of CP typically being tenths of a percent. Comparisons with integrated data
obtained earlier indicated that the sign of the CP for a given source
usually remained constant on time scales of decades. In addition, it became
clear that CP could occasionally be detected in compact components in the
inner VLBI jets.

CP measurements of AGN have recently made another major step forward
with the publication of first-epoch CP results for the 133 sources in the
MOJAVE sample of radio-bright AGN, which are regularly monitored by the
VLBA at 15~GHz (Homan \& Lister 2006).  Of these
sources, 34 displayed CP at the positions of their total-intensity peaks
(nearly always the VLBI core) at a level of $2\sigma$ or greater, with the
typical degrees of
CP being a few tenths of a percent. Comparison with 20 different optical,
radio and intrinsic parameters of the AGN yielded virtually no evidence
for any correlation between the degree of CP and these other parameters.
Another exciting result of the first-epoch MOJAVE CP measurements was
that 5 of the sample sources clearly displayed appreciable CP in their
VLBI {\em jets}, well resolved from the core, indicating that a CP-generation
mechanism capable of operating effectively in optically thin regions was
involved.

We consider here the hypothesis that the parsec-scale CP is generated via
Faraday conversion in helical jet {\bf B} fields in these AGN,
focusing on the {\em sign} of the observed CP, which has not been 
investigated in previous studies. We first examine the sign of CP generated
by Faraday conversion (Section~2), then turn to the special case of 
conversion in a helical magnetic-field geometry (Section~3). In Section~3,
we also describe our approach for using multi-frequency VLBI polarization 
data to determine the expected CP sign for particular AGN in a simple
helical jet B-field model.  Sections~4 and 5 discuss the data used for 
our analysis and our results: new detections of CP and transverse jet 
Faraday-rotation gradients in several AGN, and our remarkable result 
that the CP signs expected in our simple model agree 
with the observed signs in all eight AGN for which the comparison was
possible, provided their longitudinal {\bf B}-field components 
correspond to South poles. The robustness and implications of this  
result are discussed in Section~6, and our conclusions summarized in Section~7.

\section{The Sign of Circular Polarization Generated by Conversion}

Various possible mechanisms for the generation of CP in relativistic
astrophysical jets are discussed, for example, by Macquart \& Melrose (2000),
Wardle \& Homan (2002, 2003), Beckert \& Falcke (2002) and Ensslin (2003).
Incoherent synchrotron radiation produces a very small amount of CP
at frequencies of several to tens of gigahertz (Legg \& Westfold 1968), and
the observed degrees of CP are high enough to make this mechanism implausible.
A much more likely
mechanism is the Faraday conversion of linear-to-circular polarization during
propogation through a
magnetised plasma (Jones \& O'Dell 1977; Jones 1988); although this mechanism
has been studied fairly extensively theoretically and demonstrated to be
plausible, the action of Faraday conversion has not been directly
demonstrated observationally.

In order for Faraday conversion to operate, the observed linear polarization
electric ({\bf E}) vector must have non-zero components both parallel and
perpendicular to the {\bf B} field in the conversion region projected onto
the sky, {\bf B$_{conv}$}.
We will describe this electric vector using components parallel
to {\bf B$_{conv}$}, {\bf E$_{\|}$}, and orthogonal to {\bf
B$_{conv}$}, {\bf E$_{\perp}$}.
${\bf E}_{\|}$ excites oscillations
of free charges in the plasma, while ${\bf E}_{\perp}$ cannot,
since the charges are not free to move perpendicular
to {\bf B$_{conv}$}. This leads to a delay between
${\bf E}_{\|}$ and ${\bf E}_{\perp}$, manifest as the introduction
of a small amount of CP; the sign of the CP depends on the relative phase
of ${\bf E}_{\|}$ and ${\bf E}_{\perp}$.
Unlike Faraday rotation, Faraday conversion does
not depend on the sign of the free charges involved. A full mathematical
description of Faraday conversion is given by Jones \& O'Dell (1977); we
will be concerned here only with a qualitative understanding of the
origin of the {\em sign} of CP generated by Faraday conversion in a simple
{\bf B}-field geometry.

\begin{figure}
\centering
\includegraphics[width=0.90\textwidth]{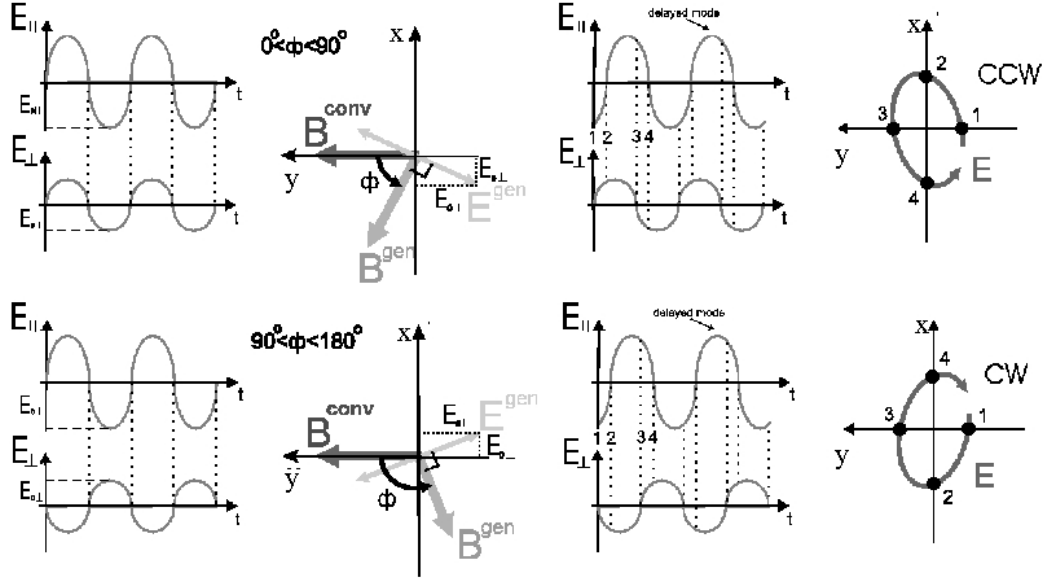}
\label{fig:conv_sign1}
\caption{Partial conversion of LP from synchrotron radiation
generated in a background field {\bf B$_{gen}$} to CP
in a foreground field {\bf B$_{conv}$} for the case when the angle 
$\Phi$ between the two fields projected onto the sky
is between $0^{\circ}$ and $90^{\circ}$ (top) and between $90^{\circ}$ and
$180^{\circ}$ (bottom). From left to right, the plots show the
relationship between the oscillations of
{\bf E$_{\|}$} and {\bf E$_{\perp}$} of the synchrotron polarization {\bf E}
vector parallel and perpendicular to {\bf B$_{conv}$}; the
relationship between {\bf B}$_{conv}$, {\bf B}$_{gen}$ and {\bf E}$_{gen}$
on the sky; and  the resulting CP that arises when {\bf E$_{\|}$} is
delayed with respect to {\bf E$_{\perp}$}. The points labeled 1, 2, 3, 4
refer to successive times. The total {\bf E} vector rotates in the
CCW direction ($+$ CP, or right-circular polarization) 
when $0^{\circ} > \Phi > 90^{\circ}$ and in the
CW direction ($-$ CP, or left-circular polarization) when 
$90^{\circ} > \Phi > 180^{\circ}$.}
\end{figure}

To demonstrate the sign of the CP that arises due to Faraday conversion, 
we consider a foreground (conversion) field {\bf B$_{conv}$}, which
purely for convenience of illustration (to avoid our diagrams from becoming
overly cluttered) we take to lie in the $y$ direction
on the sky (Fig.~1), in a right-handed ``observer's'' coordinate system in which 
North (the $x$ direction) is upward and East (the $y$ direction) is 
to the left. We suppose that LP is generated in 
an optically thin background
synchrotron source, and that the conversion delays correspond to phase
shifts of less than $90^{\circ}$. In this case, the key parameter
determining the sign of the resulting CP turns out to be the angle $\Phi$
between the synchrotron {\bf B} field, {\bf B$_{gen}$}, and {\bf B$_{conv}$},
projected onto the plane of the sky, measured in the usual sense from 
North through East, as shown (Fig.~1).  The observed linear polarization 
{\bf E} vector
will be orthogonal to {\bf B$_{gen}$}, since the synchrotron emission
region is optically thin. 

In principle, some amount of conversion is inevitable if $\Phi$
differs from $0$ ($180^{\circ}$) and $90^{\circ}$
($270^{\circ}$). The angle $\Phi$ lies in one of the four quadrants
(i) $0<\Phi<90^{\circ}$, (ii) $90^{\circ} < \Phi < 180^{\circ}$, 
(iii) $180^{\circ} < \Phi < 270^{\circ}$
or (iv) $270^{\circ} < \Phi < 360^{\circ}$.
As examples, Fig.~1 qualitatively
illustrates the connection between the angle $\Phi$, the resulting delay
between {\bf E$_{\|}$} and {\bf E$_{\perp}$} and the sign of the
resulting CP for the first two cases.
In case (i), the oscillations
of {\bf E$_{\|}$} and {\bf E$_{\perp}$} are in phase (they have the same
sign), and the tip of the vector sum of {\bf E$_{\|}$} and
{\bf E$_{\perp}$} moves in the counter-clockwise (CCW) direction on the sky,
corresponding to positive CP (right-circular polarization);
in case (ii), {\bf E$_{\|}$} and {\bf E$_{\perp}$} oscillate $180^{\circ}$ out
of phase oscillate (they have opposite signs), and the tip of the vector 
sum of {\bf E$_{\|}$} and
{\bf E$_{\perp}$} moves in the clockwise (CW) direction on the sky,
corresponding to negative CP (left-circular polarization).
Cases (iii) and (iv) are entirely analogous to cases (i) and (ii).

\section{Helical {\bf B} Fields and Conversion}


As has been mentioned by Wardle \& Homan (2002) and analysed in more detail
by Ensslin (2003), a helical jet {\bf B}-field
geometry can facilitate the conversion process -- a small fraction of the LP
from the far side of the jet relative to the observer is converted to CP
as it passes through magnetised plasma in or near the near side of the jet.
The angle $\Phi$ --- and thereby the distribution of CP ---
will vary across the jet, but in observations with limited
resolution, the overall observed CP sign should correspond to
that near the central axis, provided that the overall jet {\bf B} field
is dominated by the ordered helical component rather than a chaotic
component.  The observed CP sign and distribution
will also be affected by the viewing angle; however, we are viewing
the jets of radio-bright AGN at angles close to
$90^{\circ}$ in the jet rest frame ($\simeq 1/\Gamma$ in the observer's
frame, where $\Gamma$ is the bulk Lorentz factor for the
jet; e.g., Lyutikov, Pariev \& Gabuzda 2005; Cohen et al. 2007), 
so that this effect should not radically change the observed
picture.  In the absence of other propagation effects, such as Faraday
rotation of the plane of polarization, the sign of the
generated CP is
determined by the pitch angle $\psi$ (the angle between {\bf B} and
the helix axis projected onto the sky) and the helicity of the jet
{\bf B} field.
Table~\ref{tab:signs} summarizes the CP sign obtained for conversion in
a simple hollow helical field with pitch angle $\psi$ and a specified helicity.
These pitch angles correspond to the
case when the longitudinal {\bf B}-field component points along the jet;
if this component points opposite to the jet, the effective pitch angle 
will be greater
than $90^{\circ}$, and the CP sign for a pitch angle of $180-\psi$
will be the same as that for $\psi$ with the same helicity.
We will return to this question below.

\begin{table}
\caption{CP Signs for conversion in helical fields} \centering
\label{tab:signs}
\begin{tabular}{cccc}
\hline
Pitch-angle & Helicity & $\Phi$ range & CP Sign \\ \hline
$0 < \psi < 45^{\circ}$ & Right & $0 < \Phi < 90^{\circ}$ &$+$ \\
$45 < \psi < 90^{\circ}$ & Right & $90 < \Phi < 180^{\circ}$ &$-$\\
$45 < \psi < 90^{\circ}$ & Left & $180 < \Phi < 270^{\circ}$ &$+$ \\
$0 < \psi < 45^{\circ}$ & Left & $270 < \Phi < 360^{\circ}$ &$-$\\
\hline
\end{tabular}
\end{table}

Thus, if we are able to determine both the helicity
and pitch-angle regime for a helical {\bf B} field
associated with a particular AGN jet, the sign of the CP generated by
Faraday conversion in this helical field can be predicted.

\subsection{Estimating the {\bf B} Field Pitch Angle}

The approximate pitch angle of
a helical {\bf B} field associated with an AGN jet is
indicated by the dominant LP structure.
The observed plane of polarization will be essentially
orthogonal to the synchrotron {\bf B} field in the optically thin jets:
helical fields with relatively
large/small
pitch angles will tend to be dominated by the toroidal/longitudinal
components of their helical fields. Thus,
fields with {\em large} pitch angles should tend to have
``transverse'' {\bf B} fields or a central ``spine'' of transverse field
with a ``sheath'' of longitudinal field at one or both edges of
the jet, while fields with
small pitch angles should tend to have longitudinal {\bf B} fields
throughout the jet.  Although the observed {\bf B}-field
structure is also affected by the viewing angle, recall that we are
viewing these jets roughly ``side-on'' in the jet rest frame,
so that the above tendencies should, in general, be obeyed.

\subsection{Determining the {\bf B} Field Helicity}

If a jet has a helical {\bf B} field and there is thermal plasma in
the jet or its immediate vicinity, we will observe a RM gradient across
the jet, due to the systematically changing line-of-sight (LOS) component
of the helical {\bf B} field (Blandford 1993). The observed RMs will have 
opposite
signs at opposite edges of the jet when the viewing angle is close to
$90^{\circ}$ to the jet axis in the jet rest frame (close to $1/\Gamma$
in the observer's frame), while the RM range may encompass RM values
of only one sign when the viewing angle differs appreciably from this
value.  Such transverse RM gradients have been observed for a number of
AGN (Asada et al. 2002; Gabuzda, Murray \& Cronin 2004; Zavala \&
Taylor 2005), and several more are reported here. Note that Sikora et al. (2005)
have pointed out that the observation of Faraday rotations exceeding
approximately $45^{\circ}$ in some cases indicates that the Faraday
rotation has to be external, suggesting that it is occurring in outer
layers of the jet flow rather than throughout the jet volume. 

The helicity of the jet {\bf B} field is directly related to the direction 
of the RM gradient.
However, an ambiguity arises when we try to deduce the helicity
based on an observed transverse RM gradient. As is shown in
Fig.~2, a right-handed helix with an ``inward'' (``South'') poloidal
component will display the same sense of RM gradient as a left-handed helix
with an ``outward'' (``North'') poloidal component. Nonetheless, the two cases
could, in principle, be distinguished observationally because they will give 
rise to different signs of CP, due to their different helicities.

\begin{figure*}
\centering
\includegraphics[width=0.42\textwidth]{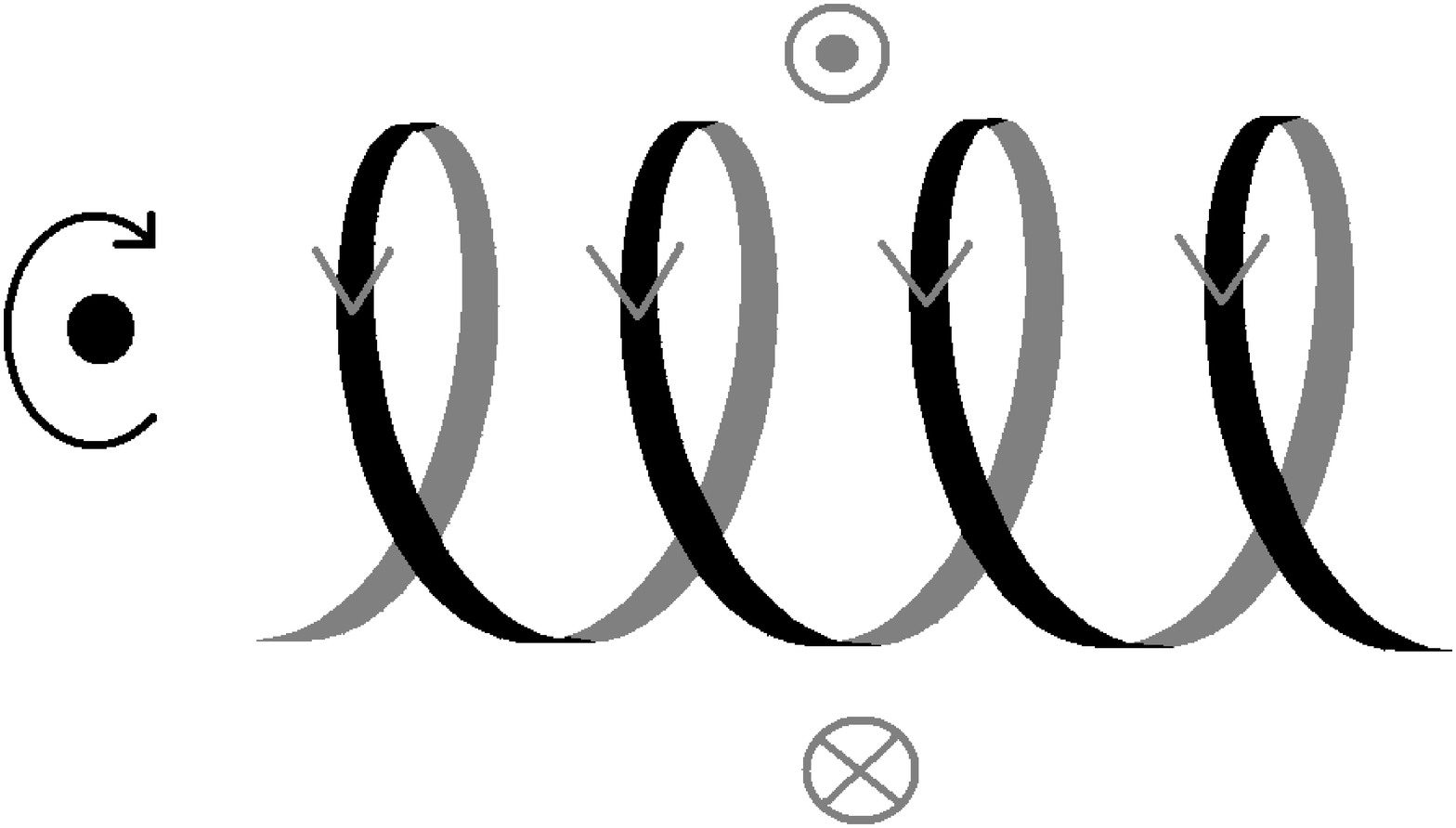}
\includegraphics[width=0.42\textwidth]{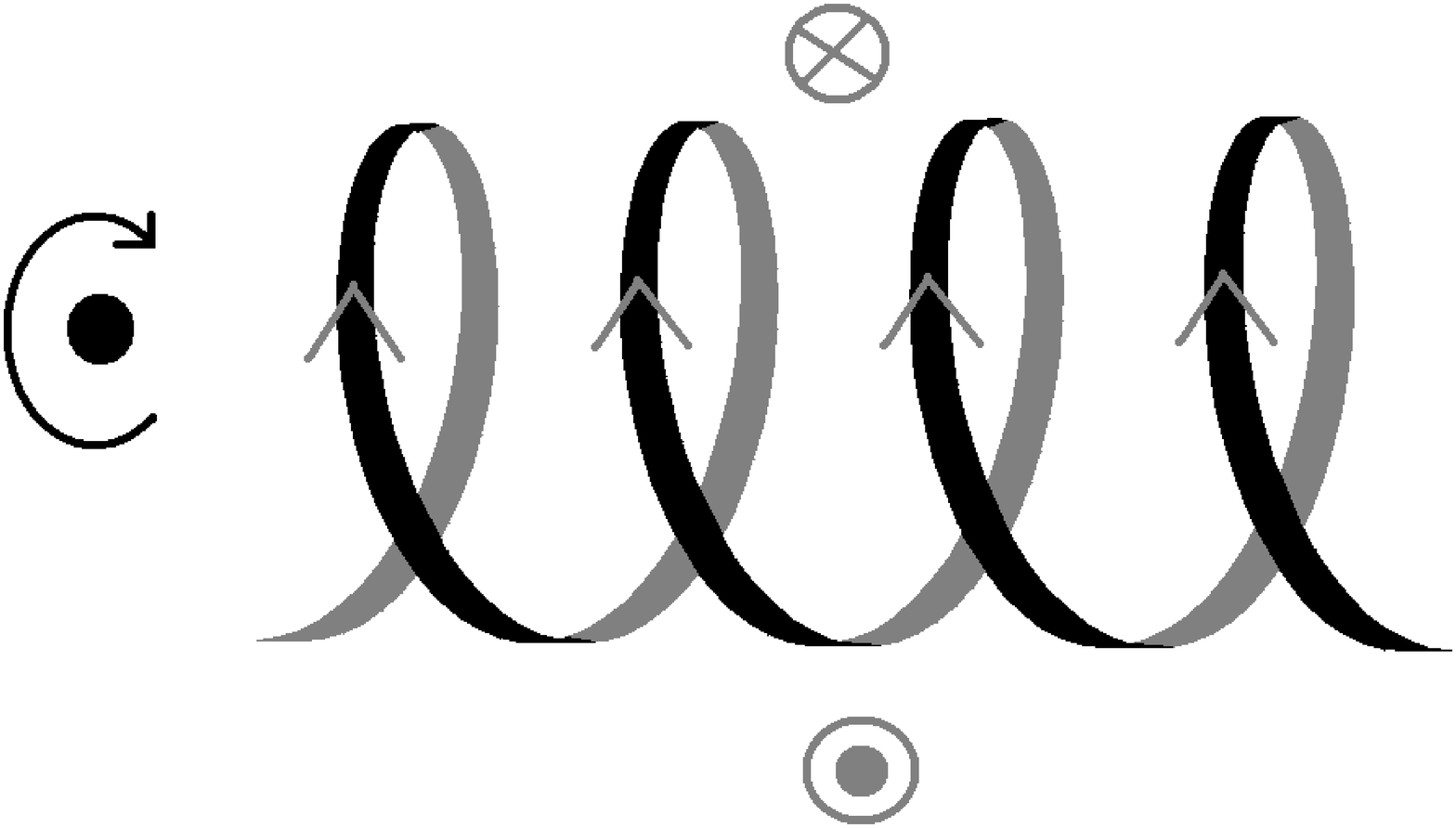}
\centering
\includegraphics[width=0.42\textwidth]{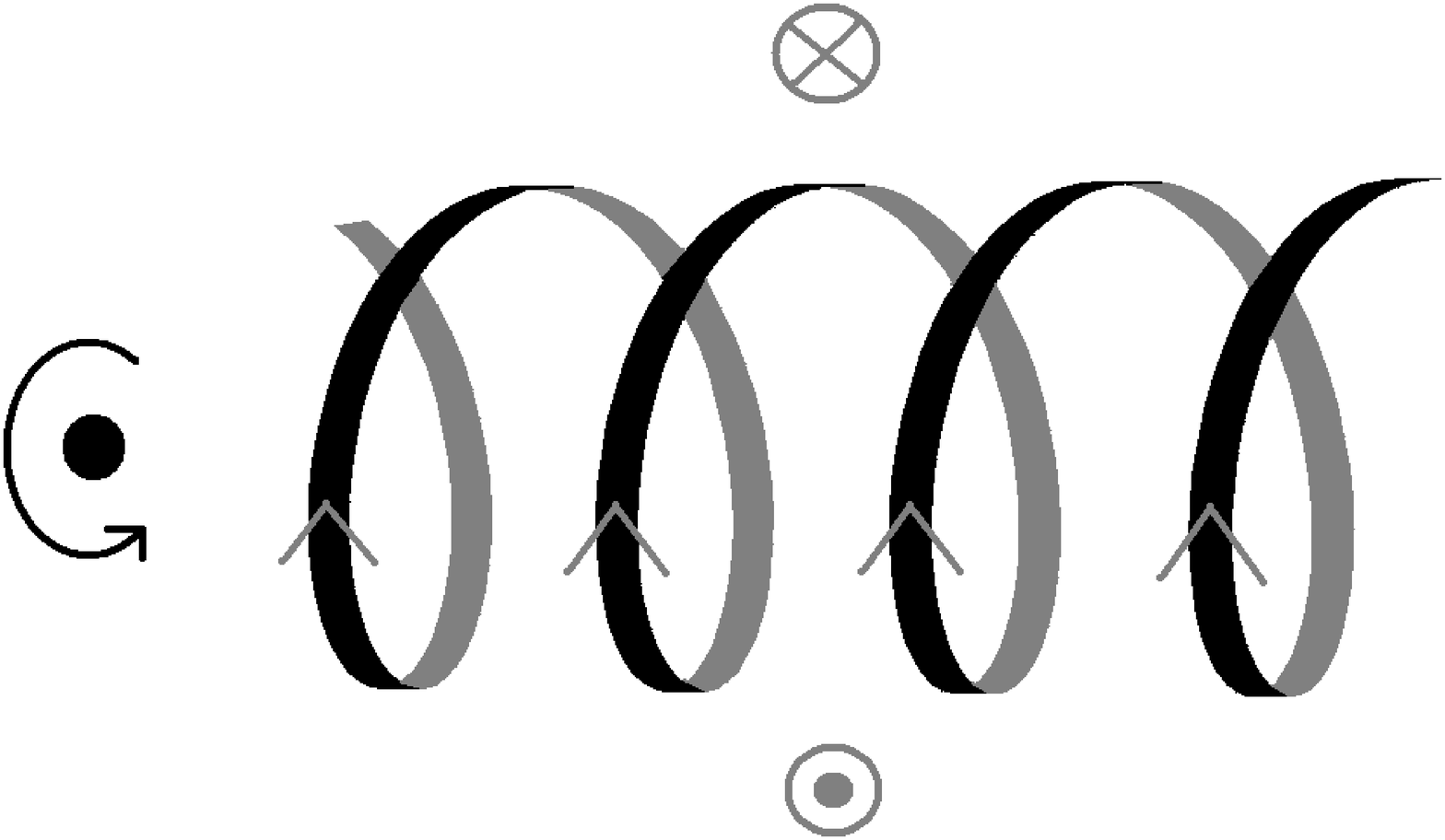}
\includegraphics[width=0.42\textwidth]{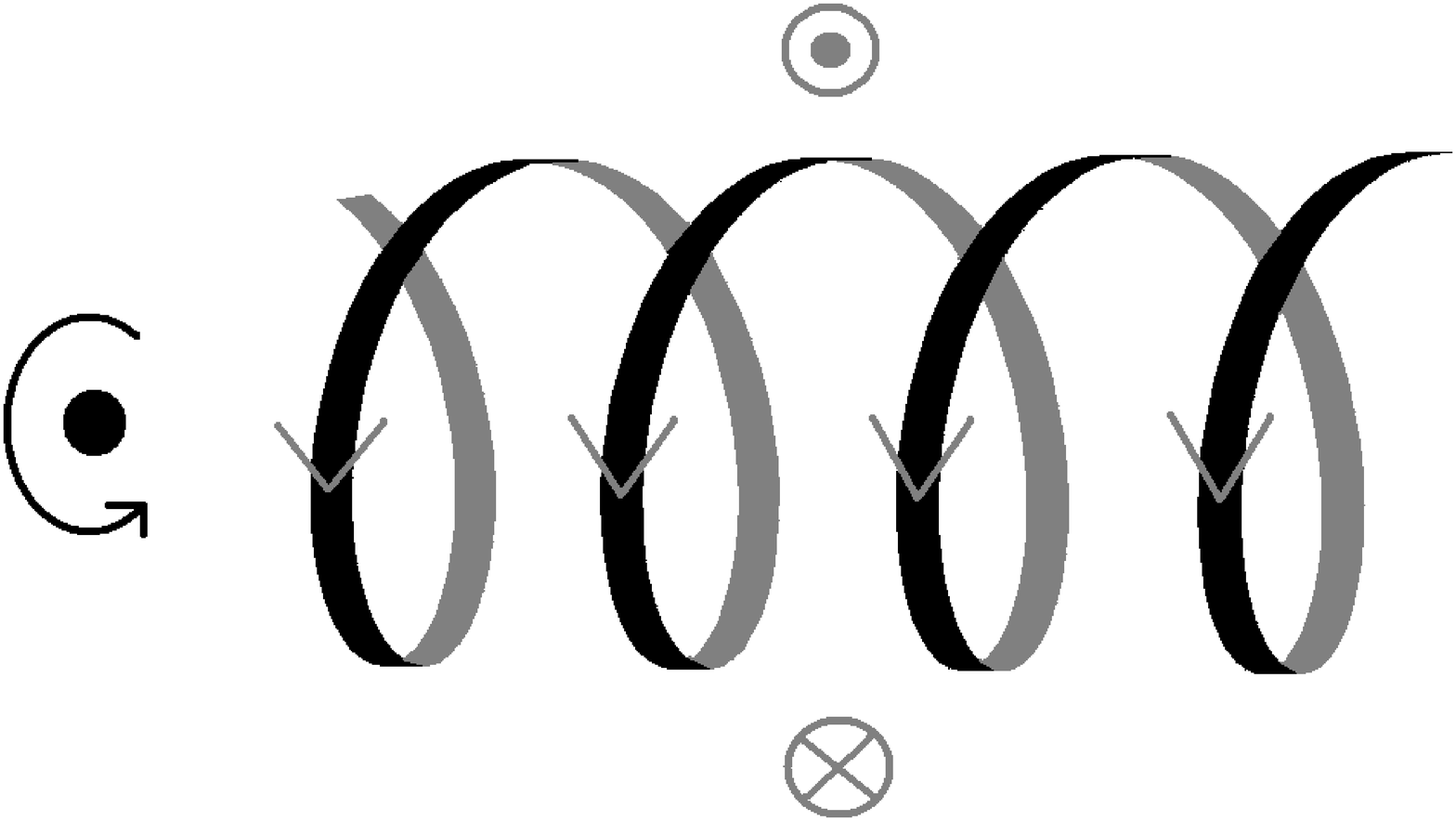}
\label{fig:sign_plots}
\caption{Diagrams showing the senses of the transverse Faraday-rotation 
gradients brought about by a jet {\bf B} field in the form of a right-handed 
(top) and left-handed (bottom) helix with poloidal component oriented away 
from (left) and toward (right) the central black hole, which is located in 
all cases to the left of the region shown. The direction of the rotation of the 
central black hole/accretion disc is also indicated to the left. A circled 
dot indicates that the toroidal component of the field is pointed toward the 
observer, 
and a circled cross that this component is pointed away from the observer.}
\end{figure*}

\section{The Data and their Reduction}

We have identified 8 AGN for which (i) parsec-scale CP has been detected,
(ii) transverse RM gradients have been detected and (iii) the LP
structure is clear enough to be reasonably certain about the correct
inferred jet {\bf B}-field structure: 0735+178, 1156+295, 3C273, 3C279,
1641+399 (3C345), 1749+096, 2230+114 and 2251+158 (3C454.3).

\begin{figure*}
\centering
\includegraphics[width=.94\textwidth]{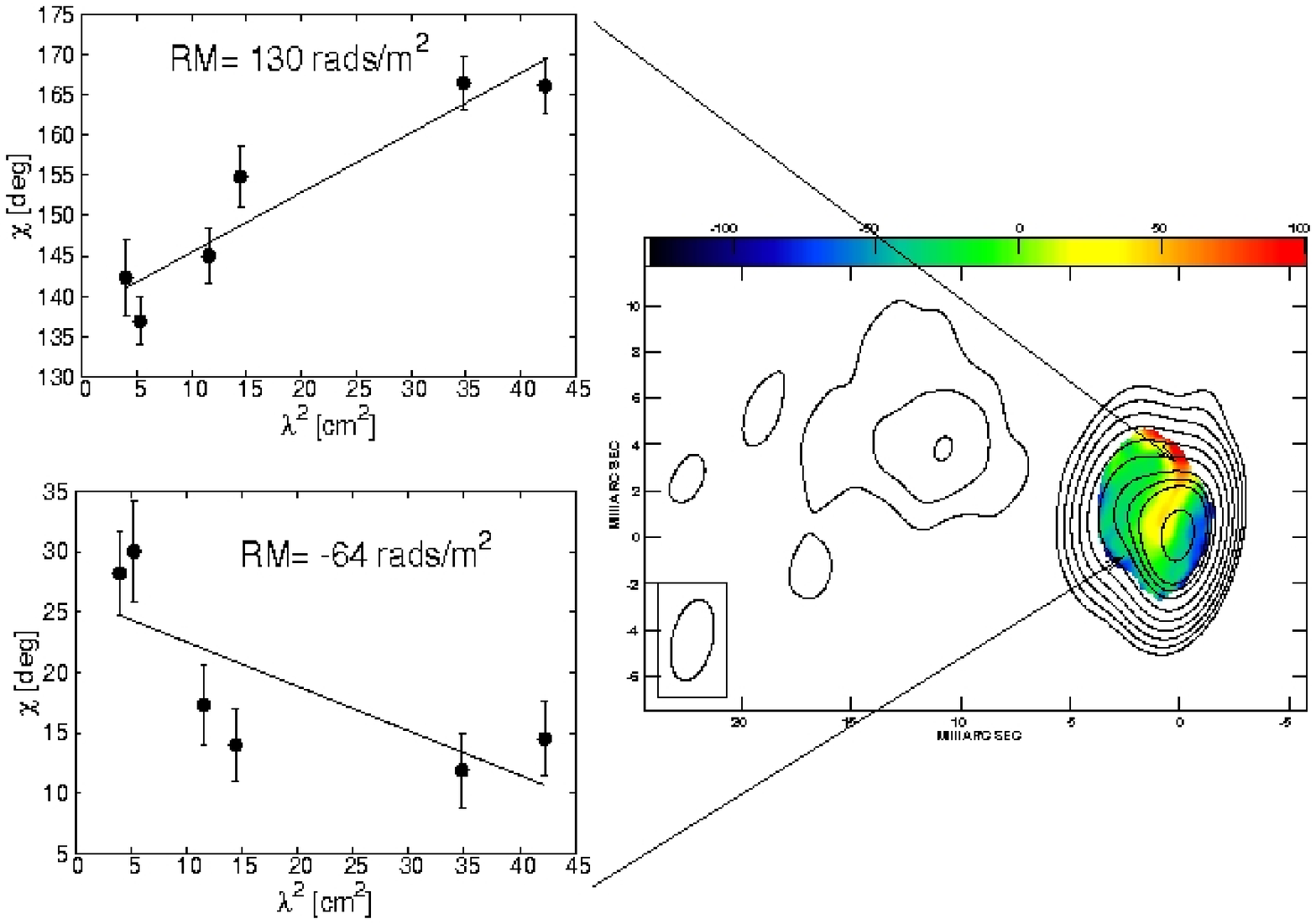}
\centering
\vspace*{0.1cm}
\centering
\includegraphics[width=.94\textwidth]{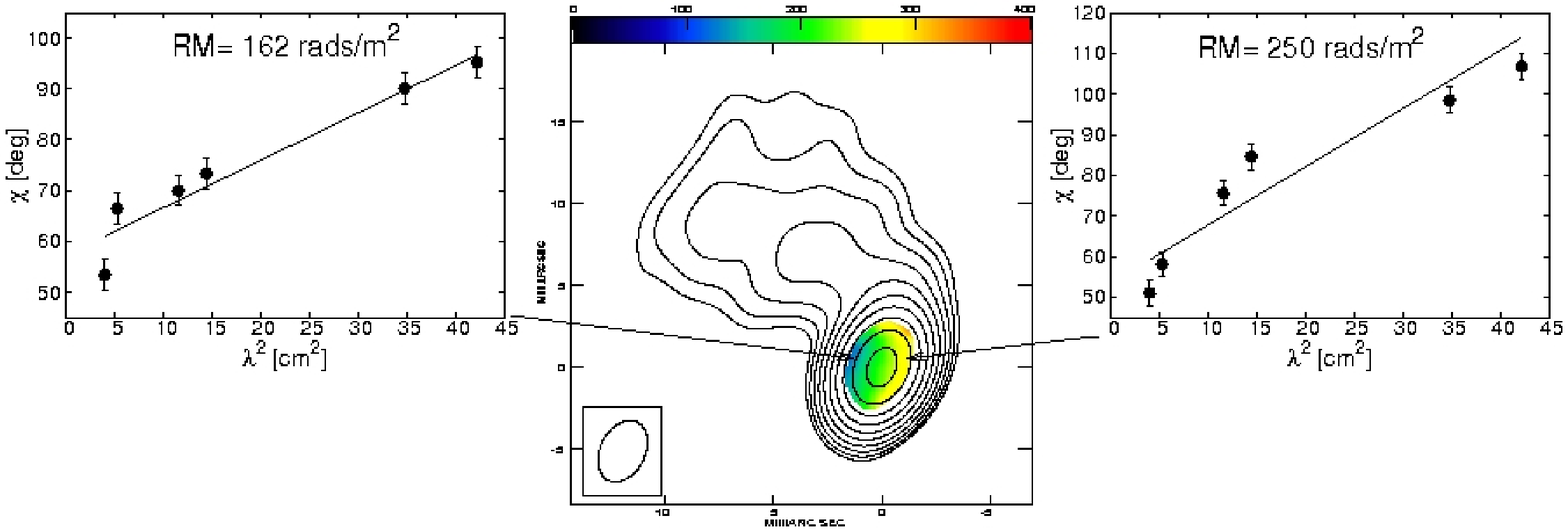}
\centering
\vspace*{0.1cm}
\includegraphics[width=.94\textwidth]{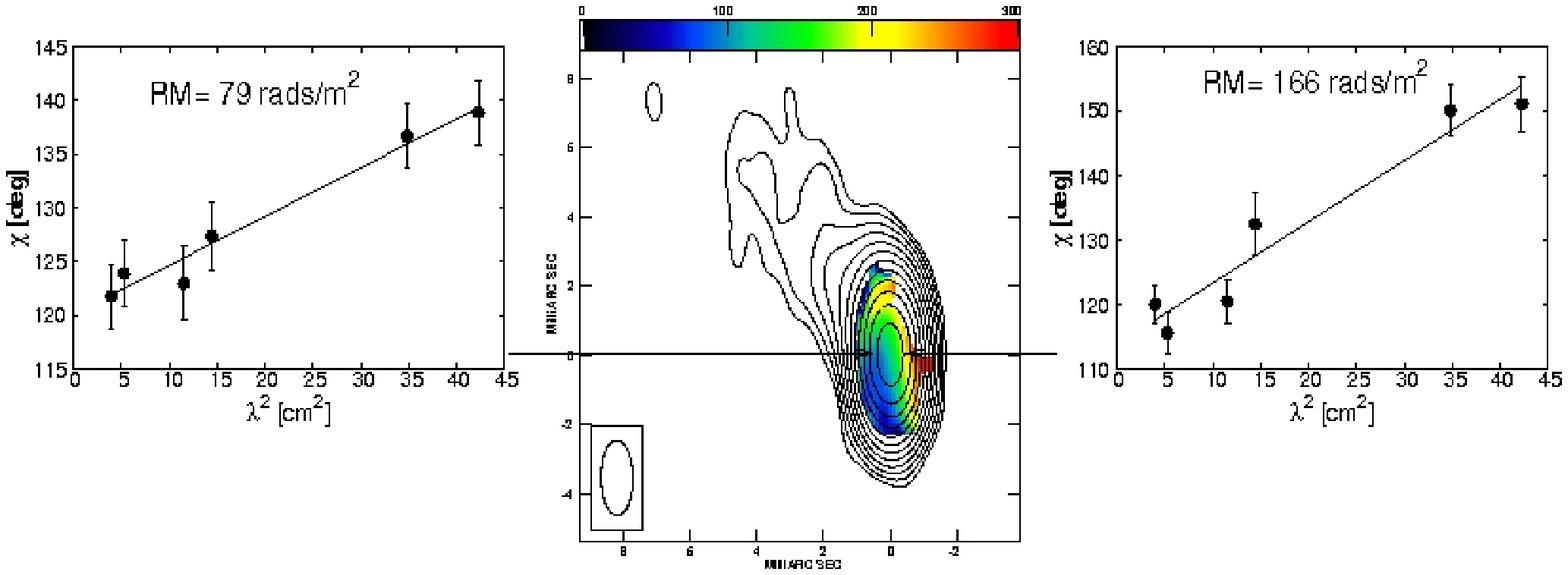}
\label{fig:rmgrads}
\caption{VLBA $I$ maps of 0735+178 (top; 4.6~GHz; peak 0.62~Jy/beam;
bottom contour 2.0~mJy/beam), 1156+295 (middle; 5.1~GHz; peak 1.32~Jy/beam;
bottom contour 1.6~mJy/beam) and 1749+096 (bottom; 7.9~GHz; peak
3.30~Jy/beam; bottom contour 1.0~mJy/beam), with the parsec-scale 
RM distributions superimposed. In all cases, North is upward, East is to
the left and the contours increase in steps of a factor of two.
The accompanying graphs show plots of $\chi$ (in degrees) vs. 
$\lambda^2$ (in cm$^2$) for
locations on opposite sides of the VLBI jets, with $\chi$ errors
of $1\sigma$.  The convolving beams are shown in the bottom right-hand
corner of the maps. The maximum allowed uncertainty in the RM values per pixel
were 30~rad/m$^2$ for 0735+178, 10~rad/m$^2$ for 1156+295 and 25~rad/m$^2$
for 1749+096.
}
\end{figure*}

Five of these 8 AGN appear in this list due
to the new detections of transverse RM gradients (0735+178, 1156+295
and 1749+096) and CP (3C345 and 2231+114) reported here.
In all cases, the target sources were observed in a ``snapshot'' mode,
with typically 8--10 scans of each object spread out over the time the
source was visible with most or all of the VLBA antennas.  The data
reduction and imaging were done in the AIPS package using standard
techniques.  The instrumental polarizations (``D-terms'') were determined
using the AIPS task LPCAL, solving simultaneously for the source polarization.

The calibration of the absolute electric vector position angles (EVPAs)
was determined using integrated polarization measurements obtained using
the Very Large Array near in time to our observations
(www.aoc.nrao.edu/~smyers/calibration/), by requiring
that the EVPA for the total VLBI
polarization of compact sources coincide with the EVPAs for their VLA cores.
We estimate that our overall EVPA calibration is accurate to
within about $3^{\circ}$. The rms noises in the maps of the Stokes parameters
$Q$ and $U$ were used to estimate the uncertainties
in the VLBA polarization angles $\chi$ in each pixel of the $\chi$ maps.
Note that any errors in the EVPA calibration affect all polarization
angles in all regions of all the sources in the same way, and so could
hinder the detection of Faraday-rotation gradients, but will not
lead to the presence of spurious Faraday-rotation gradients in the source
structure.

\subsection{Faraday-Rotation Observations}

The polarization observations used for our Faraday-rotation measurements
were obtained on 22 August 2003 (1156+295), 22 March 2004 (1749+096) and
10 September 2004 (0735+178) at 15.1, 12.9, 8.6, 7.9, 5.1 and 4.6~GHz using
the NRAO VLBA.
The polarization D-terms were derived from observations
of DA193 (August 2003), 0235+164 (March 2004) and 1732+389 (September 2004).

We made maps of the distribution of the Stokes parameters $Q$ and $U$
with matched resolutions corresponding to the 4.6~GHz or 8.1~GHz beam
(indicated in a corner of the map figures).
The $Q$ and $U$ maps were than used to construct the distributions of the
polarized flux ($p = \sqrt{Q^2 + U^2}$) and polarization angle
($\chi = \frac{1}{2}\arctan \frac{U}{Q}$), as well as accompanying ``noise
maps,'' using the AIPS task COMB. The formal uncertainties written in the
output $\chi$ noise maps were calculated in COMB using the rms noise levels on
the input $Q$ and $U$ maps.

\begin{table*}
\caption{15.3~GHz CP Results for 1 November 2004} \centering
\label{tab:cp_bg152a}
\begin{tabular}{lccccccc}
\hline
Source   & Alias &  $I$ peak (Jy) & $m_c$ (\%)& $\sigma$ \\
0048$-$097&      & 0.55 & $< 0.12$ & $-$ \\ 
0256+075 &       & 0.22 & $< 0.18$ & $-$ \\ 
0804+499 &       & 0.56 & $< 0.07$ & $-$ \\ 
0906+430 &       & 0.66 & $< 0.08$ & $-$ \\
1156+295 &       & 0.48 & $< 0.08$ & $-$ \\
1633+382 &       & 2.63 & $-0.23\pm 0.06$ & 3.8 \\
1641+399 & 3C345 & 1.98 & $+0.15\pm 0.06$ & 2.5 \\
2134+004 &       & 0.91 & $< 0.13$ & $-$ \\
2230+114 & CTA102& 1.06 & $-0.47\pm 0.12$ & 3.9 \\
\hline
\end{tabular}
\end{table*}

\begin{table*}
\caption{AGN Jets with Transverse RM Gradients and CP} \centering
\label{tab:results}
\begin{tabular}{lcccccccccc}
\hline
Source   & Dominant jet & Pitch-angle & Helicity & Expected & Helicity & Expected & Observed & Implied & CP & RM \\
& {\bf B} field${\dagger}$&  regime    &   if N {\bf B}$_{pol}$  & CP sign & if S {\bf B}$_{pol}$ & CP sign & CP (\%) & {\bf B}$_{pol}$ & Ref & Ref \\ \hline
0735+178 & $\perp$ & Large    &  Left    &  $+$   &  Right & $-$ & $-0.30\pm 0.11$ &S  & 1 & *\\
1156+295 & SS      & Large    &  Left    &  $+$   &  Right & $-$ & $-0.27\pm 0.09$ &S & 1 & *\\
3C273    & $\|$    & Small    &  Right   &  $+$   &  Left  & $-$ & $-0.45\pm 0.09$ &S & 1 & 3, 4\\
3C279    & $\perp$ & Large    &  Left    &  $-$   &  Right & $+$ & $+0.30\pm 0.08$ &S & 1 & 5\\
3C345    & $\|$    & Small    &  Left    &  $-$   &  Right & $+$ & $+0.17\pm 0.10$ &S & * & 6\\
1749+096 & $\perp$ & Large    &  Left    &  $+$   &  Right & $-$ & $-0.32\pm 0.13$ &S & 2 & *\\
2230+114 & $\|$    & Small    &  Right   &  $+$   &  Left  & $-$ & $-0.48\pm 0.11$ &S & * & 7\\
2251+158 & SS      & Large    &  Right   &  $-$   &  Left  & $+$ & $+0.23\pm 0.10$ &S & 1 & 8\\
\hline
\multicolumn{11}{l}{$^{\dagger}$$\perp$ = transverse, $\|$ = longitudinal,
SS = spine+sheath}\\
\multicolumn{11}{l}{1 = Homan \& Lister 2006; 2 = Vitrishchak \& Gabuzda 2007;
3 = Asada et al. 2002; 4 = Zavala \&}\\
\multicolumn{11}{l}{Taylor 2005; 5 = Zavala \& Taylor 2004; 6 = Taylor 1998;
7 = Taylor 2000; 8 = Zavala \& Taylor 2003; * = this paper}
\end{tabular}
\end{table*}

The RM maps were constructed as is described by Gabuzda, Murray \&
Cronin (2004), using $\chi$ maps with matched beam sizes
at the six frequencies.
We removed the contributions of the known integrated (predominantly
Galactic, i.e., foreground; Pushkarev 2001) Faraday rotations at each
frequency before making the RM maps, so that any
remaining rotation measures should be due to thermal plasma in the
vicinity of the AGN. The AIPS task RM has the option of blanking output
pixels when the uncertainty in the RM exceeds a specified
value; the maximum allowed RM uncertainties are indicated in the caption
to Fig.~3.

When determining the $\chi$ values for individual regions
in the VLBI jet, we found the mean within a $3\times 3$~pixel
($0.3\times 0.3$~mas) area at the corresponding
location in the map; the polarization angles were assigned errors
equal to the rms deviation for this mean value and the estimated
EVPA calibration uncertainty of $3^{\circ}$ added in quadrature.

\begin{figure*}
\centering
\includegraphics[width=0.43\textwidth]{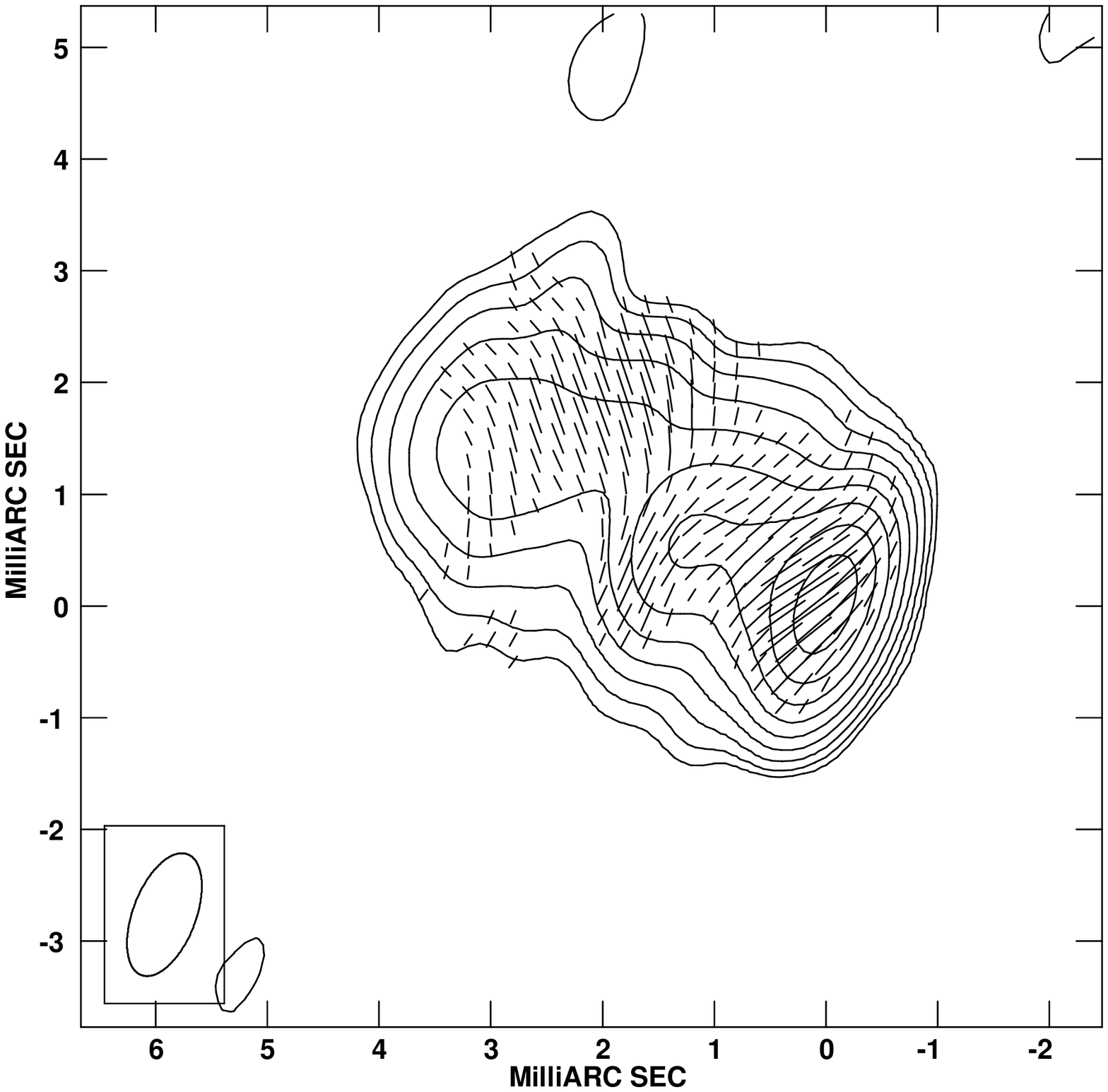}
\includegraphics[height=10.5cm]{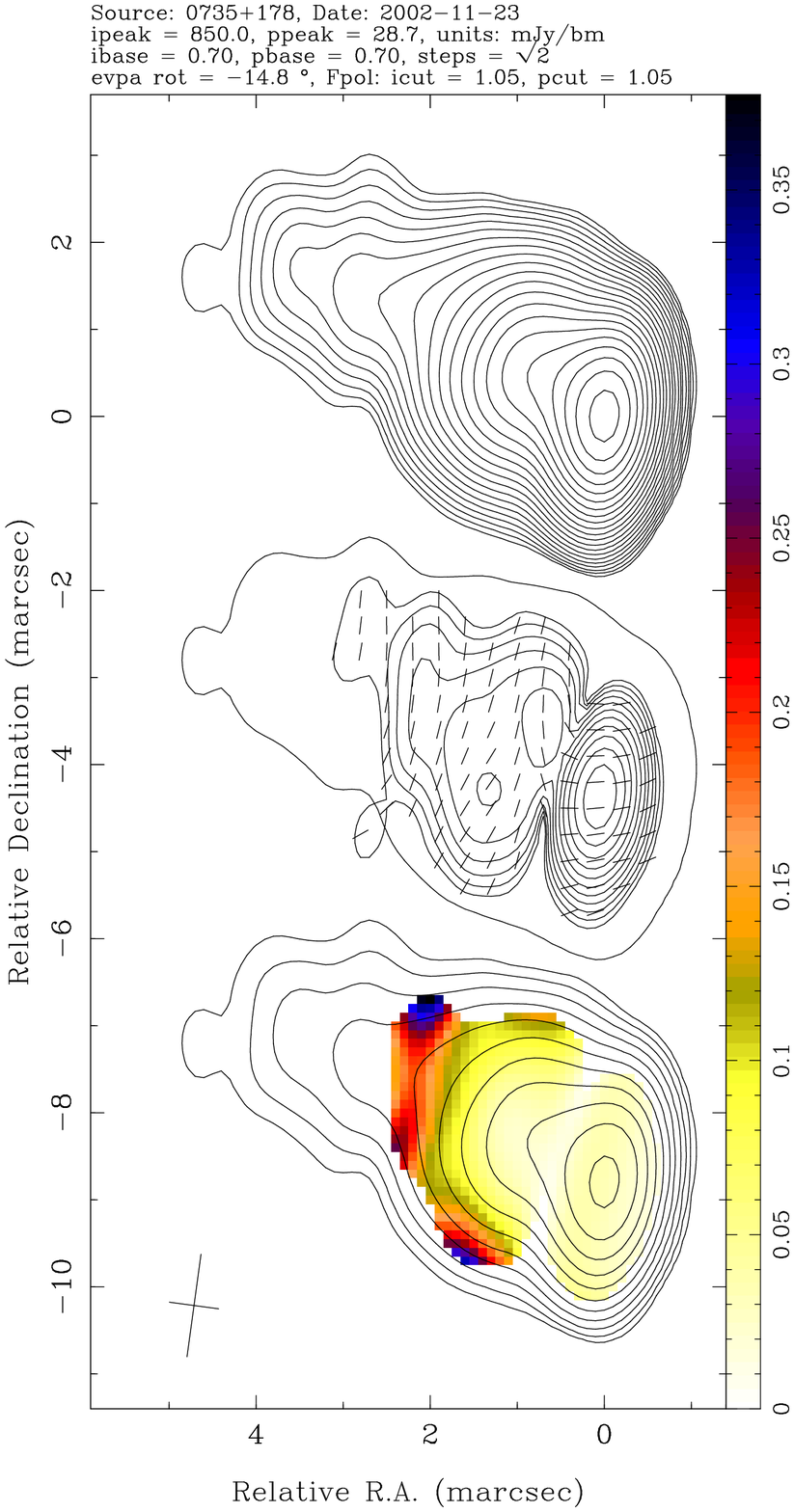}
\includegraphics[width=0.43\textwidth]{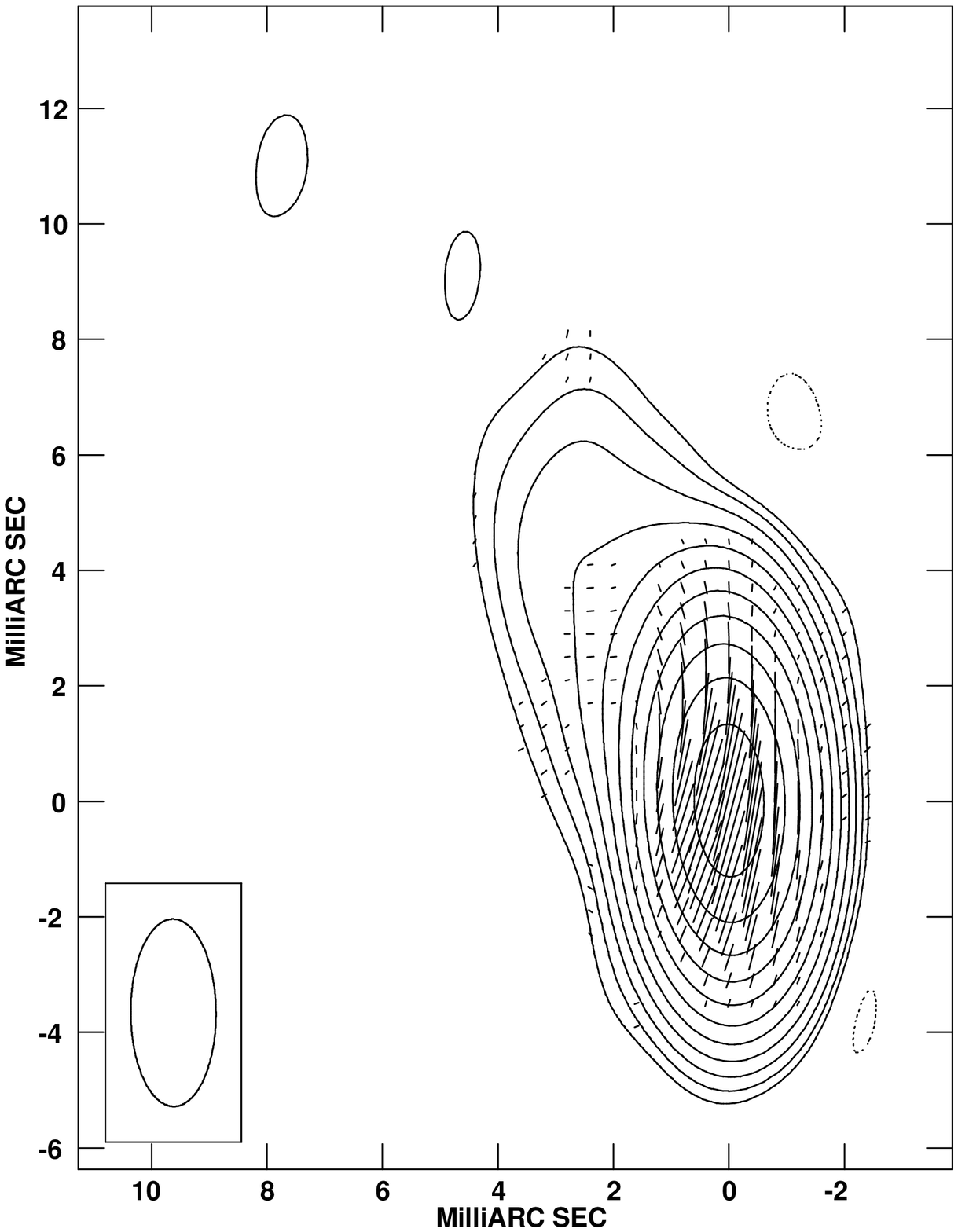}
\label{fig:maps1}
\caption{Top left: VLBA $I$ map of 0735+178 with polarization
sticks superimposed from our data obtained on 10 September 2004 at 15.1~GHz
(peak 0.45~Jy/beam; bottom contour 1.1~mJy/beam; contour step of
two). Top right: VLBA $I$ map (peak 0.85~Jy/beam; bottom contour 
0.70~mJy/beam; contour step of $\sqrt{2}$), polarized flux map with 
polarization sticks superimposed and $I$ map with the distribution of 
the degree of polarization superimposed at the first MOJAVE epoch on 
23 November 2002 at 15.4~GHz. The predominance of a transverse {\bf B} 
field in the core region at the MOJAVE epoch is clearly visible. Bottom: 
VLBA $I$ map of 1749+096 with polarization 
sticks superimposed from our 5.1-GHz data obtained on 22 March 2004 
(peak 2.21~Jy/beam; bottom contour 1.4~mJy/beam; contour step of two); the 
polarization structure
shown in the first-epoch MOJAVE data obtained on 31 May 2002 is similar. 
The convolving beams are shown in the bottom right-hand corner of the maps.}
\end{figure*}

\begin{figure*}
\centering
\includegraphics[width=0.40\textwidth]{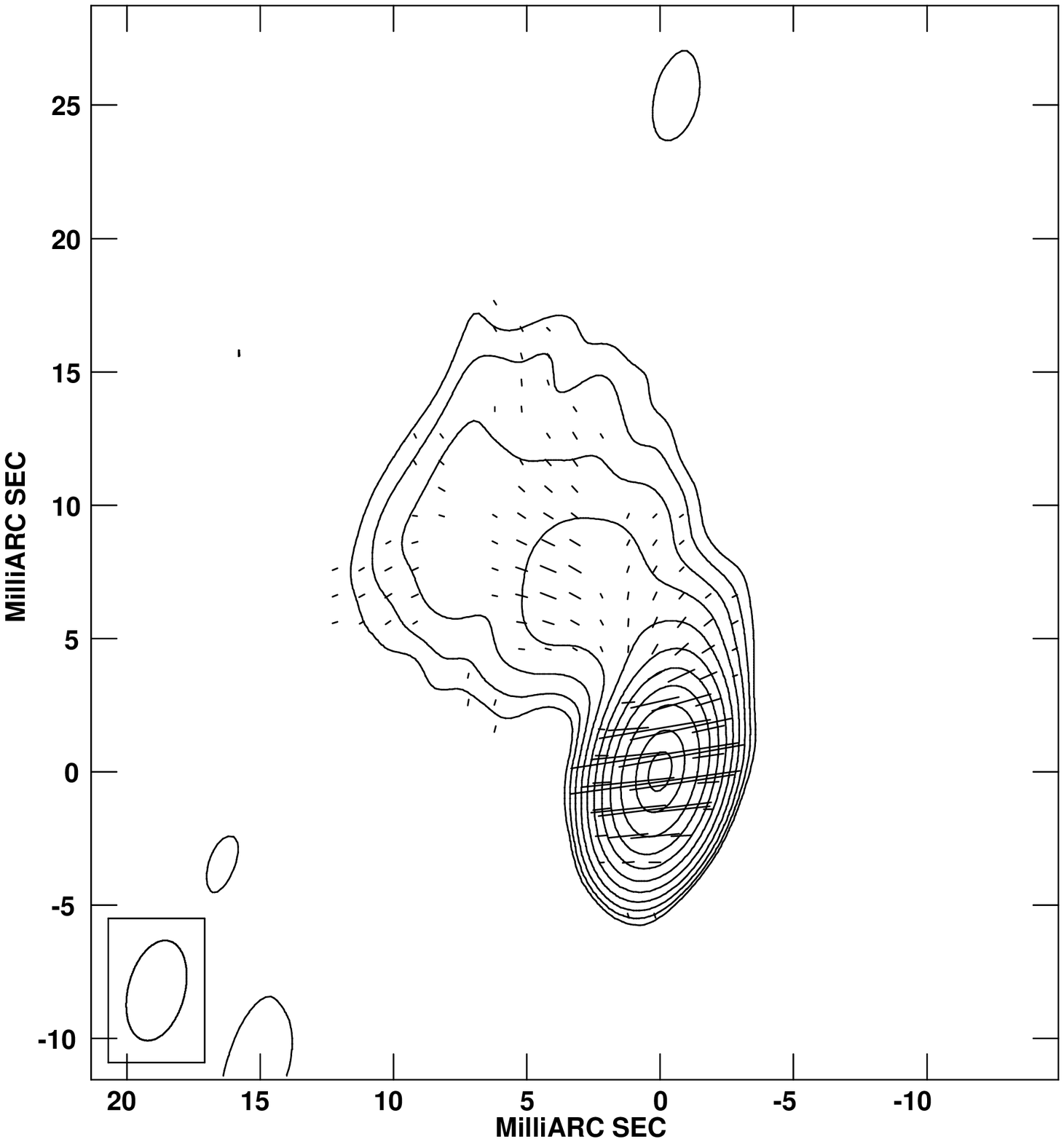}
\includegraphics[width=0.40\textwidth]{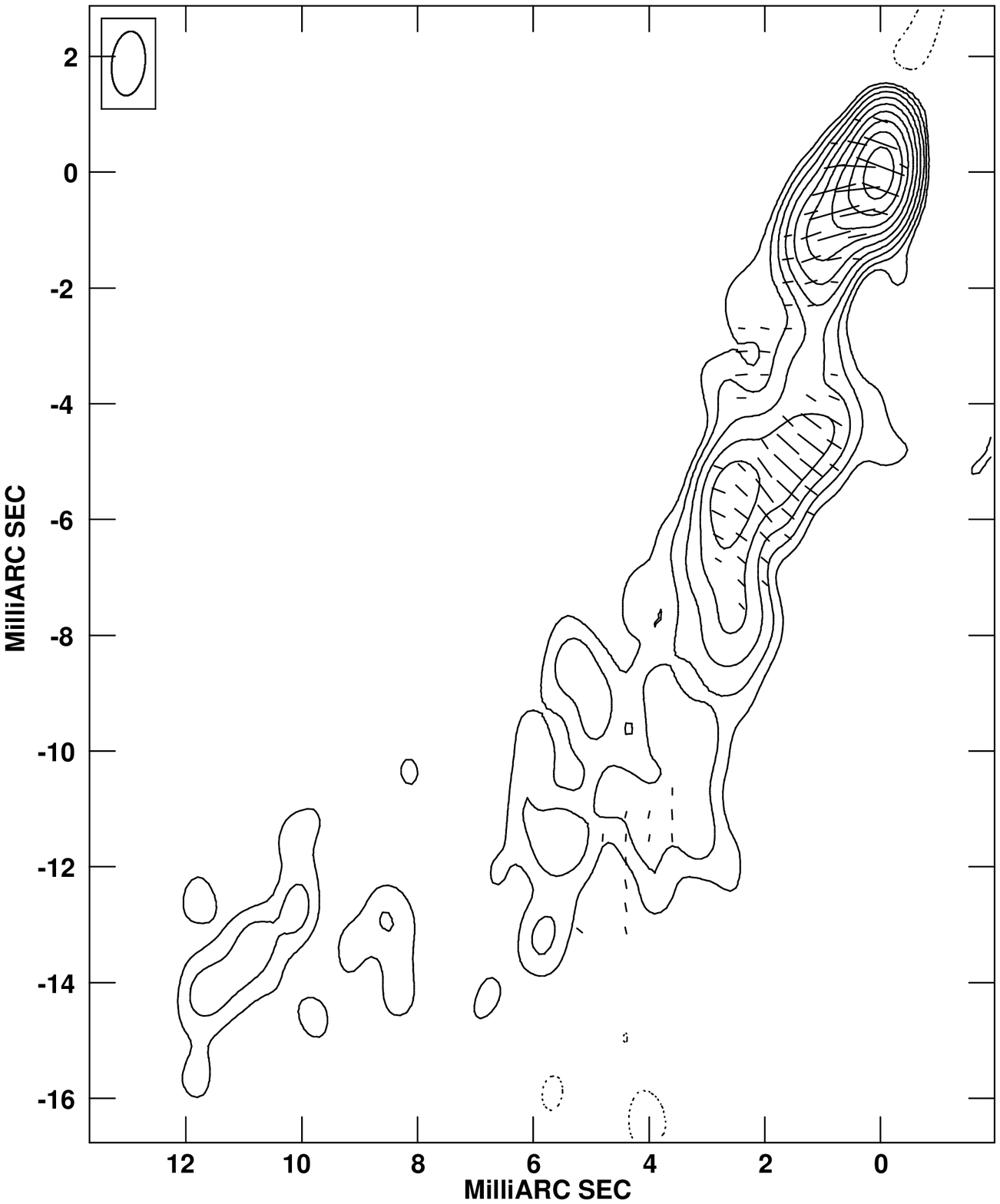}
\centering
\includegraphics[width=0.50\textwidth, angle=-90]{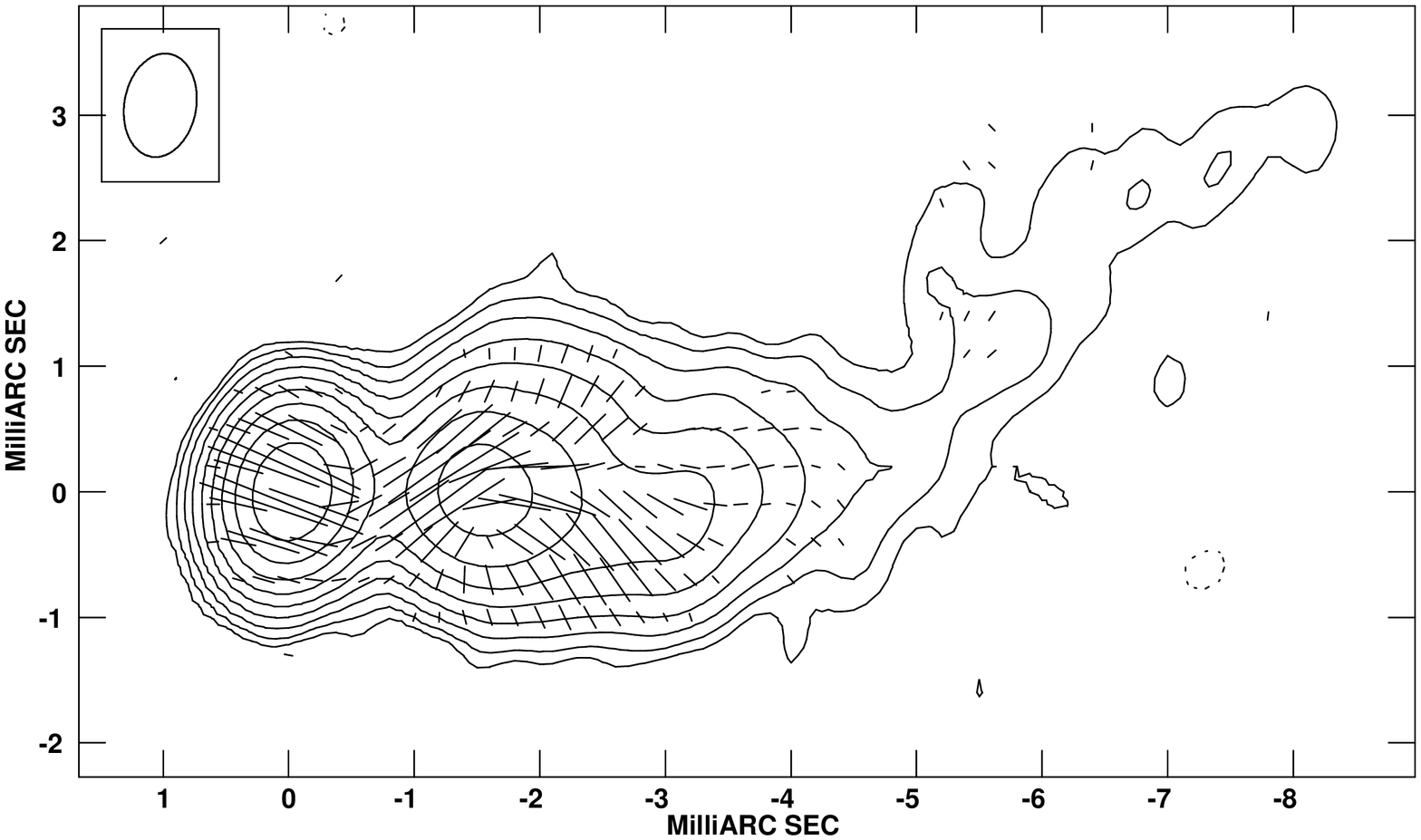}
\label{fig:maps2}
\caption{VLBA $I$ maps with polarization sticks superimposed
at 5.1~GHz at epoch 22 August 2003 for 1156+295 (top left; peak 1.32~mJy/beam;
bottom contour 1.6~mJy/beam), and at 15.3~GHz at epoch 1 November 2004 for
2230+114 (top right; peak 1.06~mJy/beam; bottom contour 2.7~mJy/beam) and
1641+399 (bottom; peak 1.98~mJy/beam; bottom contour 4.2~mJy/beam). In
all cases, the contours increase in steps of a factor of two.
The convolving beams are shown in some corner of the maps.}
\end{figure*}

\subsection{VLBA CP Observations}

The VLBA observations of 10 AGN used in our CP analysis were obtained
at 15.3~GHz  on 1 November 2004.  The polarization D-terms were derived
from observations of 1156+295.

The circular-polarization calibration of the gains was done using the
well established gain-transfer techique of Homan and Wardle (1999),
described in detail both in that paper and by Vitrishchak \& Gabuzda (2007).
CP was detected in three of the ten AGN observed: 1633+283, which was
found to have a CP of $(-0.23\pm 0.06)\%$, consistent with the value of 
$(-0.39\pm 0.09)\%$
reported by Homan and Lister (2006), and 1641+399 [$(+0.15\pm 0.06)\%$] and 
2230+114 [$(-0.47\pm 0.12)\%$], for which we report the first CP detections 
here. A table of all the CP measurements for our 1 November 2004 15.3-GHz
data are given in Table~\ref{tab:cp_bg152a}. The $1\sigma$ uncertainties in the CP
values were determined in the same way as is described by Homan et al. (2001),
including contributions for uncertainty (i) in the smoothed antenna gains,
(ii) due to real CP in the calibrators, and (iii) due to
random scan-to-scan gain variations. In the case of non-detections, the
presented upper limits are the corresponding $1\sigma$ uncertainties.
The uncertainty-estimation procedure used
is relatively conservative, and we are in the course of investigating a
Monte Carlo approach to estimating the CP uncertainties, similar to that 
used by Homan \& Lister (2006).

\section{Results}

Fig.~3 shows the VLBI $I$ images of 0735+178, 1156+295
and 1749+096 with color images of their parsec-scale RM distributions
superposed. The convolving beams used in each case are indicated in the
lower right-hand corner of the figure. The RM gradients across the VLBI
jets are visible by eye. The accompanying plots show the observed polarization
position angles, $\chi$, as a function of the observing wavelength squared,
$\lambda^2$, for individual locations in the VLBI jets, with $1\sigma$
errors in $\chi$.

The observed total-intensity and linear-polarization structures for each
of the 8 objects listed above at the first MOJAVE epoch are presented
by Lister \& Homan (2005); additional images for other epochs are available
at the site http://www.physics.purdue.edu/astro/MOJAVE/.
We present the total-intensity and linear-polarization structures for the
5 of these 8 objects for which we have reported new transverse
RM gradients or CP measurements in Figs.~4 and 5.

Let us briefly discuss the {\bf B}-field structures we infer from the
linear-polarization structure of each of the 8 sources considered here
(for the purpose of determining the approximate pitch angles of the helical
{\bf B} fields associated with their jets). Note that the Faraday rotations
in the central regions of the jets typically produce rotations of no more
than $5-10^{\circ}$ at the wavelengths considered here, so that this does 
not drastically affect our ability to draw conclusions about the jet 
{\bf B}-field structures.

{\em 0735+178.} This BL~Lac object has been analysed in detail in a number of
VLBI studies (Gabuzda et al. 1994; Gabuzda, G\'omez \& Agudo 2001; G\'omez
et al. 2001; Agudo et al. 2006). In some time intervals, the VLBI jet has 
appeared relatively straight,
while in others, it has shown an unusual ``zig--zag'' structure, with two
sharp bends in the inner jet. The MOJAVE maps and our own image in
Fig.~4 may correspond to a transition from the zig--zag to a
straighter jet structure (Agudo et al. 2006), but the origin of this
transition (e.g., changes in the projected bends on the plane of the sky,
changes in the physical structure of the jet itself) is not entirely clear.
Our image seems to indicate a predominantly longitudinal jet 
{\bf B} field (the polarization vectors lie perpendicular to the jet); however,
the previous studies cited above have indicated a clear tendency for the
dominant {\bf B} field in the inner jet to be {\em orthogonal} to the jet.
Fig.~4 also shows the first-epoch MOJAVE images of 0735+178,
which, like our own image, seem to display an extended region of longitudinal
field in the outer jet, but also clearly show a region in the core/inner jet
which has a transverse {\bf B} field. Overall, given that earlier
images have shown clear evidence for a predominant transverse jet {\bf B}
field in the inner jet and there have been dramatic changes in the observed
jet structure, we suggest that the extended regions of longitudinal field
indicated by the maps in Fig.~4 may reflect interactions between
the jet and the surrounding medium, rather than the intrinsic pitch angle of
the helical jet {\bf B} field. Thus, although this is a somewhat problematic
case, we have taken the predominant intrinsic jet {\bf B} field in 0735+178 
(i.e., that field that reflects the pitch angle of the helical field) to be
orthogonal to the jet.

{\em 1156+295.} The VLBI jet of 1156+295 initially emerges nearly due North,
then curves to the East. Our image in Fig.~5 shows a clear
spine+sheath type structure in the linear polarization, with the inferred
{\bf B} field perpendicular to the jet near the jet axis and longitudinal
to the jet near its edges, consistent with the MOJAVE polarization images.
Polarization observations of 1156+295 at 15, 8.4
and 5~GHz obtained in September 2005 show a very similar polarization
structure, and also confirm the presence of a transverse RM gradient
with the same sense (Gabuzda, Mahmud \& Kharb, in preparation).

{\em 3C273.} A number of polarization studies have been devoted to this 
well known quasar; it was the first AGN for which a transverse RM gradient
indicative of a helical {\bf B} field was reported, by Asada et al. (2002),
subsequently confirmed by Zavala \& Taylor (2005) and Attridge et al. (2005).
The dominant inferred {\bf B} field is longitudinal throughout the observed
jet, including the smallest scales probed by 43~GHz VLBA observations
(Jorstad et al. 2005).

{\em 3C279.} This object has also been included in many studies; in addition,
it is often observed as a calibrator due to its brightness, compactness and the 
relative stability of its polarization angles. The MOJAVE images show that
the dominant {\bf B} field is orthogonal to the jet throughout the observed 
jet, and this tendency is retained on the smallest scales probed by 43~GHz 
VLBA observations (Jorstad et al. 2005).

{\em 1641+399 (3C345).} A number of polarization studies have been devoted to 
this well known quasar (e.g. Ros, Zensus \& Lobanov 2000 and references 
therein).  It is one of only a few AGN for which there is firm evidence
that different jet components have been ejected in different structural
position angles, and follow different trajectories, which ultimately merge
to form the more outer VLBI jet. Although individual components with {\bf B}
fields orthogonal to the jet have been detected in the innermost jet, these
may be associated with shocks; the predominant jet {\bf B} 
beyond $1-2$~mas from the core has consistently been observed to be 
longitudinal, and we accordingly adopt this latter field as
the {\bf B} field that should be used to deduce the pitch-angle regime. 

{\em 1749+096.} This very compact BL~Lac object was included in a number
of earlier studies (e.g. Gabuzda et al. 2000 and references therein; Gabuzda
2003), and in all cases the dominant {\bf B} field has been orthogonal
to the jet. Our new 5.1~GHz image in Fig.~4 displays such an
orthogonal field in the inner part of the jet, possibly with signs of a
spine+sheath like structure appearing further from the core. The 15.4~GHz
MOJAVE image likewise clearly shows a 
region of transverse {\bf B} field in the inner jet, to the North of the core; 
the jet makes a sharp turn to the East, and the polarization sticks in this more
extended region remain aligned with the new jet direction.  A hint
of a transverse RM gradient in 1749+096 with the same sense as the one
shown in Fig.~3 is visible in the RM map of
Zavala \& Taylor (2004).

{\em 2230+114.} The behavior of this AGN is similar to that of 3C345:
close to the VLBI core, there is a compact region whose polarization is
sometimes aligned with or oblique to the jet direction, while the 
predominant {\bf B} field beyond $1-2$~mas is clearly longitudinal (e.g.
Fig.~5). Again, based on its stability in time, we adopt
the longitudinal field as the {\bf B} field that should be used to deduce
the pitch-angle regime. 

{\em 2251+158 (3C454.3).} The VLBI jet of this well-known quasar is very 
complex, but the MOJAVE polarization images display a clear spine+sheath 
structure, and the orthogonal {\bf B} field near the central ridge line 
of the jet clearly dominates at higher frequencies (Jorstad et al. 2005).

Table~\ref{tab:results} summarises the {\bf B}-field structures, pitch-angle
regimes implied by these {\bf B}-field structures [``small''~$=0^{\circ} <
\psi < 45^{\circ}$ and
``large''~$=45^{\circ} < \psi < 90^{\circ}$], 
helicities implied by the observed transverse RM gradients for the case of
North and South poloidal magnetic-field components {\bf B}$_{pol}$
and the expected CP signs for each of these two cases. This table also presents 
the observed parsec-scale CP values. {\em In all 8 AGN, the observed sign of the CP
agrees with the sign expected for our simple helical-field model for the case
of a South poloidal field.} This clearly non-random pattern of agreement
between the expected and observed CP signs represents powerful evidence that the
detected CP is generated by Faraday conversion in the helical jet
{\bf B} fields of these objects, accompanied by unexpected evidence
that the poloidal fields in the jets preferentially correspond to South polarity.

If the detected CP had no connection with the presence of helical fields,
the predicted and observed CP signs should agree and disagree in roughly
half the cases for each sense of the poloidal magnetic field. Testing this hypothesis
using the binomial probability distribution, specifying equal probabilities
for the expected and observed CP signs agreeing or disagreeing, yields
a probability of only $0.4\%$ that the perfect agreement between our predicted 
CP signs for the case of South poloidal field and the observed
CP signs came about purely by chance. 

\section{Discussion}

\subsection{Location of Detected CP in the VLBI Cores}

It is important to note that the ``core'' as the optically thick base of the
jet (Blandford \& K\"onigl 1979) is a theoretical concept, and will correspond
to the observed ``core'' only for observations with sufficient resolution; 
the observed VLBI ``core'' will actually correspond to a combination of the 
genuine optically thick core and optically thin emission from the inner jet.
Taking this into account, in our model, the fact that the detected CP usually 
coincides with
the VLBI ``core'' is essentially a selection effect. The Faraday conversion 
occurs to some extent on all scales in the jets, including on smaller scales 
than the resolution of the VLBI observations, and it is natural that the
inner jet dominates the CP signal, since this is the brightest part of the
jet. In addition, this effect may be enhanced by the fact that 
the {\bf B}-field strengths and (thermal-)plasma densities that determine the 
conversion may rise toward the active nucleus of the AGN. 

It is interesting that CP has, in fact, been
directly detected in the VLBI {\em jets} of 7 AGN (Homan \& Lister 2006,
Vitrishchak \& Gabuzda 2007), 3 of which
appear in Table~2 (3C273, 3C279, 2252+158).  Of these 7 AGN,
all but 3C84, which is clearly an unusual object (Homan \& Wardle 2004),
display extended regions of CP of a single sign.
This behaviour is fully consistent with the CP being
associated with a helical jet {\bf B} field whose  pitch angle and
helicity determine the sign of the observed CP.

\subsection{Sensitivity of Results to Viewing Angle}

We have carried out the above analysis assuming that we view most AGN jets
at angles not very different from $1/\Gamma$ ($90^{\circ}$ in the jet rest
frame). For example, the analysis of Cohen et al. (2007) suggests that the
most likely angle to the line of sight is approximately $0.6/\Gamma$ 
($\simeq 60-65^{\circ}$ in the jet rest frame for typical values of $\Gamma$). 
One natural question is how sensitive our results are to this
assumption. In other words, for how large a range of angles do we expect
to observe the CP sign predicted for a viewing angle of $\simeq 90^{\circ}$ in the
jet rest frame?

\begin{figure*}
\centering
\includegraphics[width=0.42\textwidth]{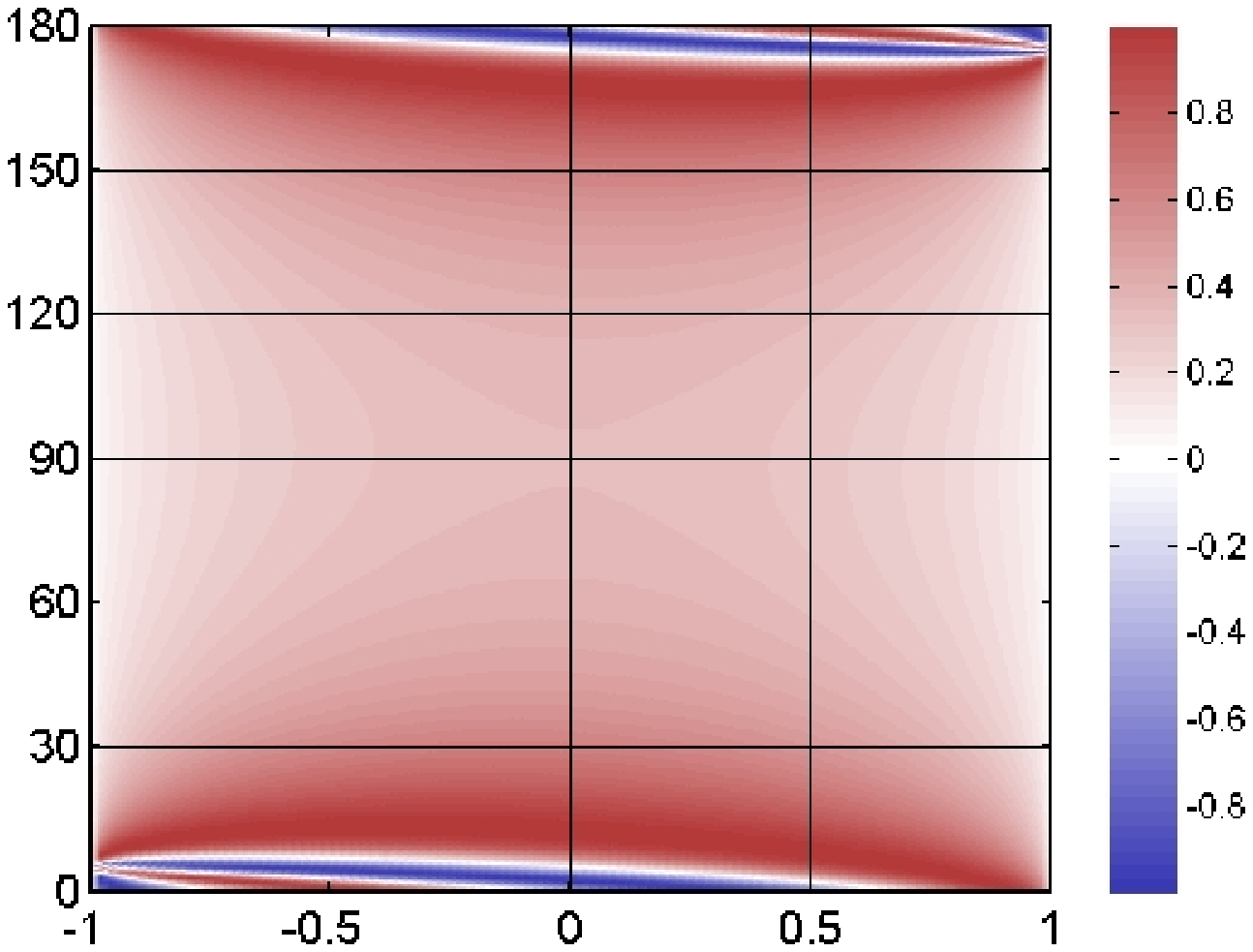}
\includegraphics[width=0.42\textwidth]{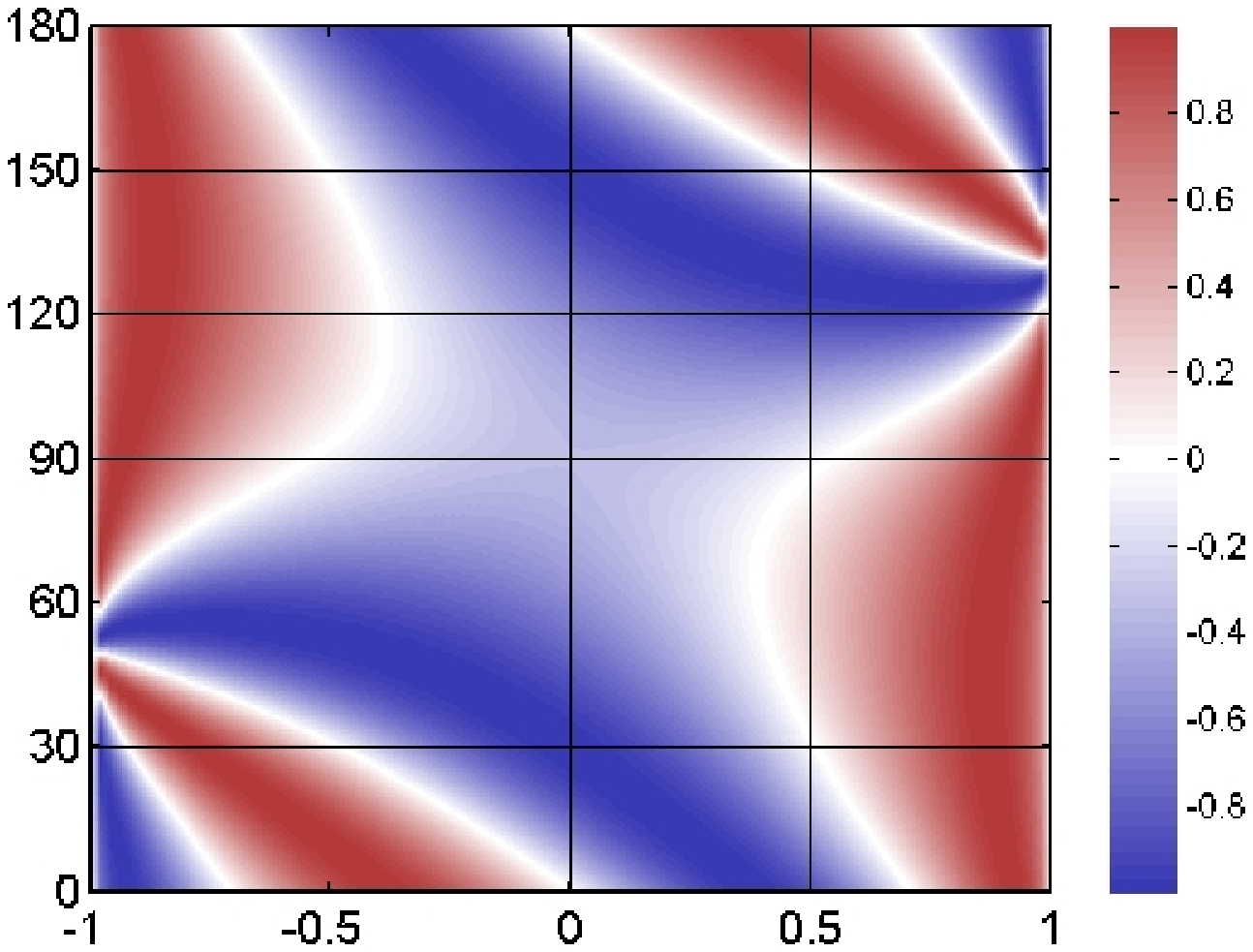}
\centering
\includegraphics[width=0.42\textwidth]{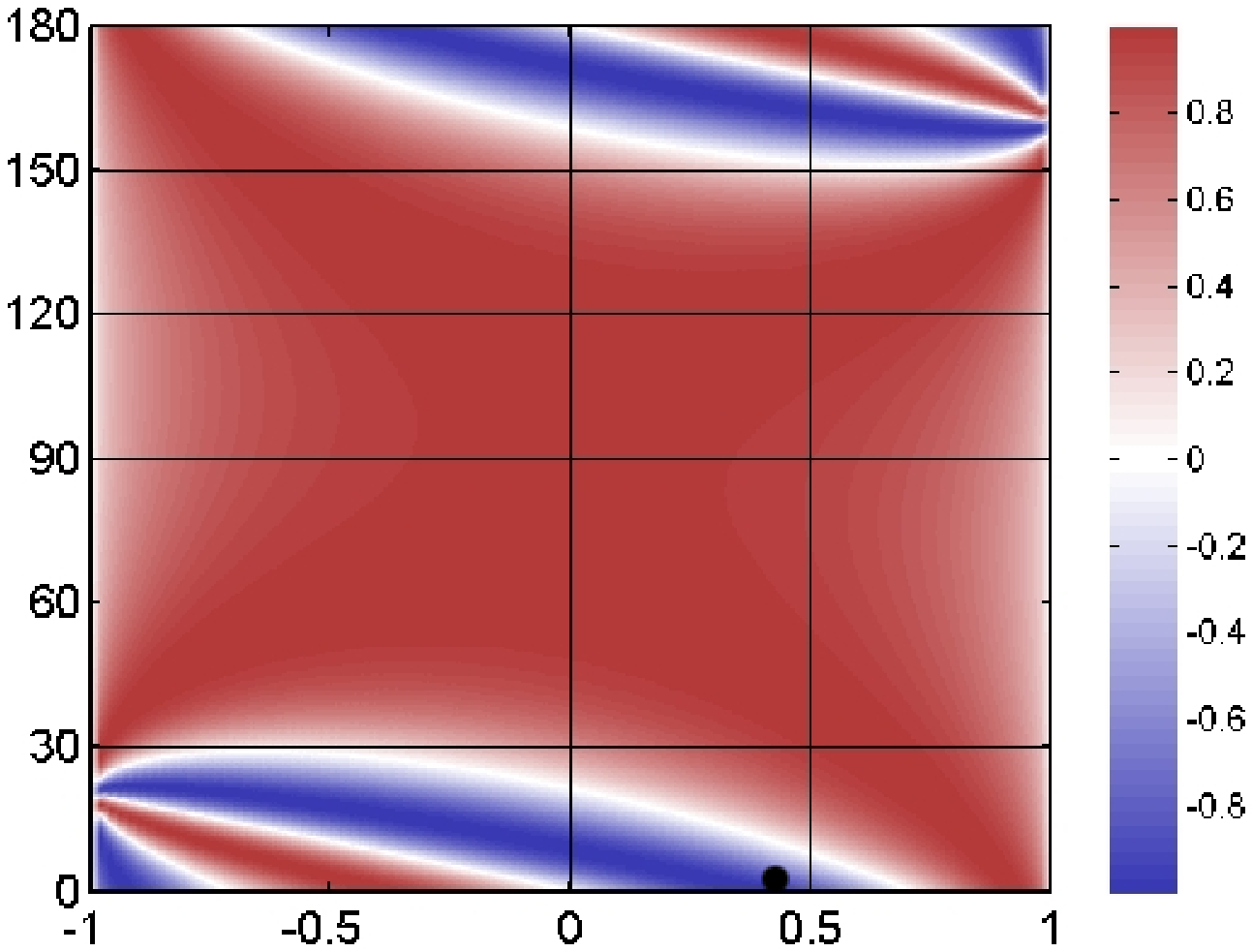}
\includegraphics[width=0.42\textwidth]{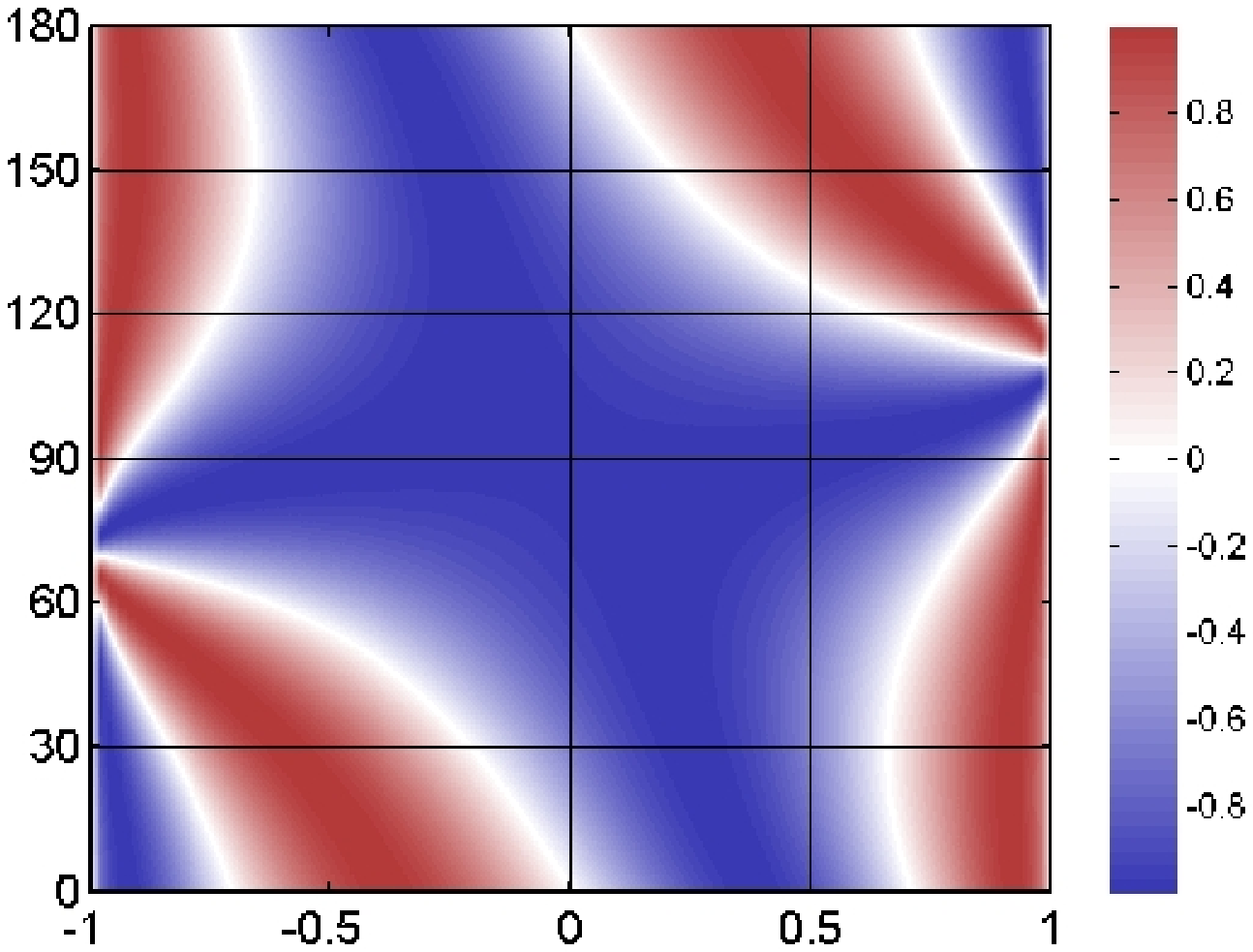}
\centering
\includegraphics[width=0.42\textwidth]{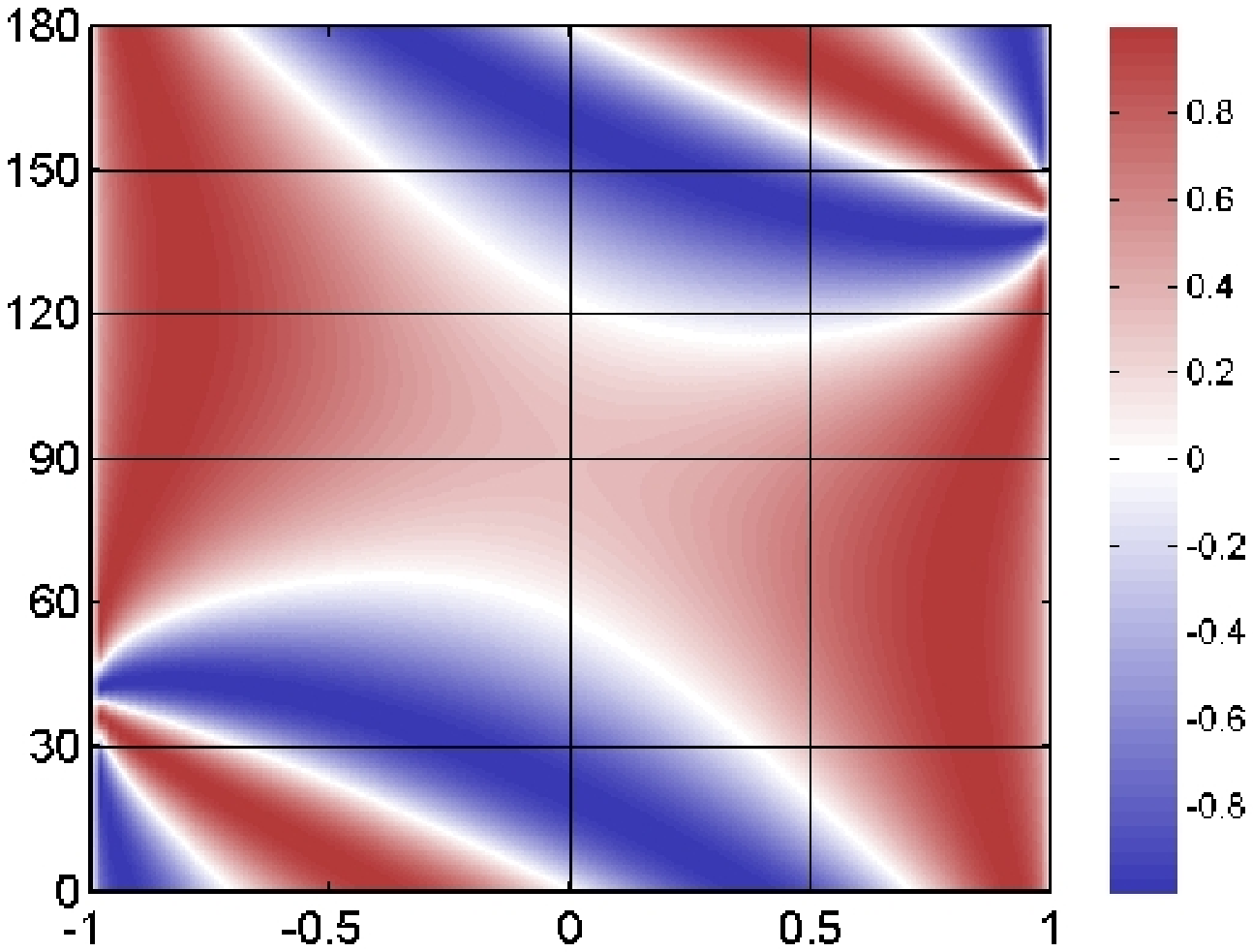}
\includegraphics[width=0.42\textwidth]{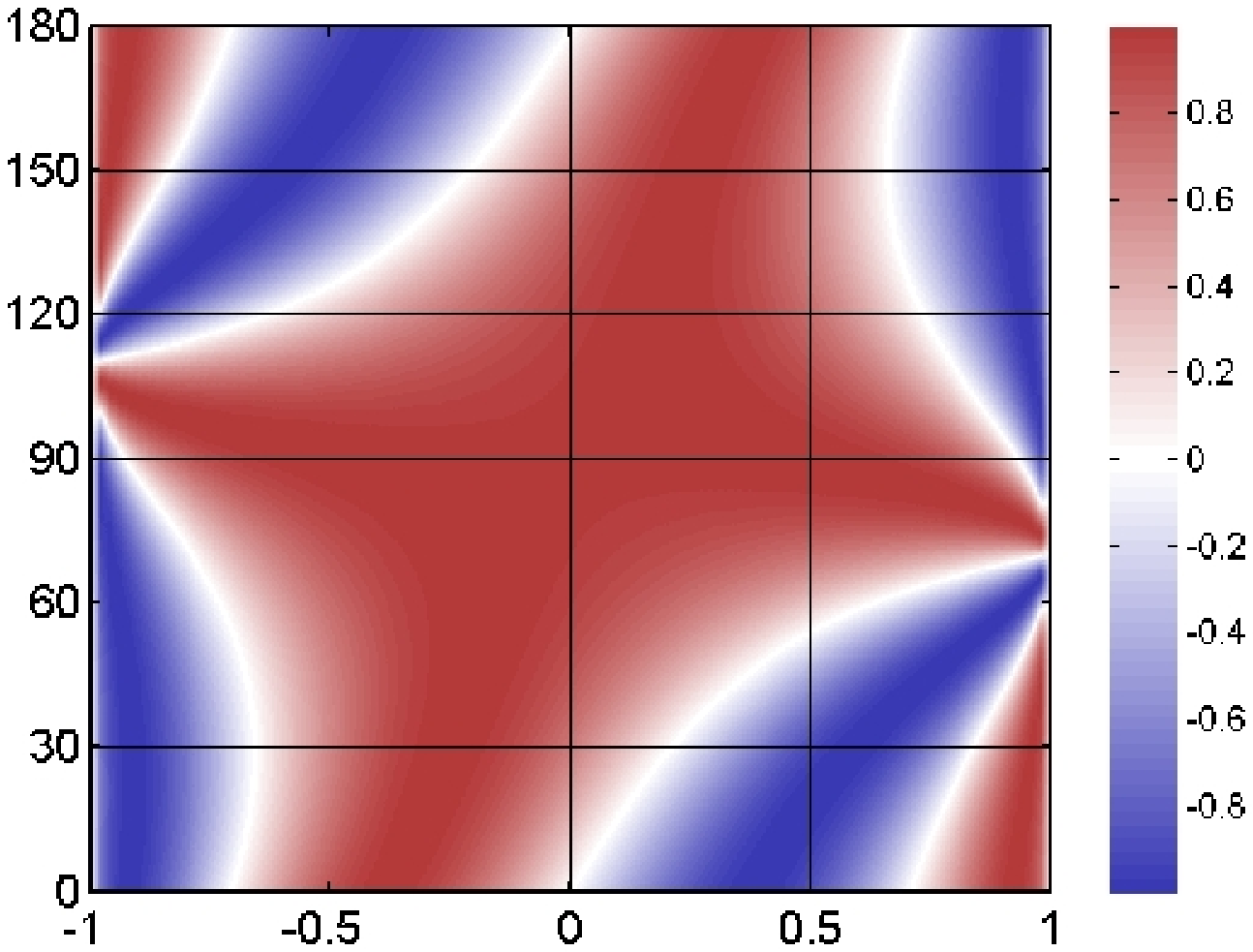}
\centering
\includegraphics[width=0.42\textwidth]{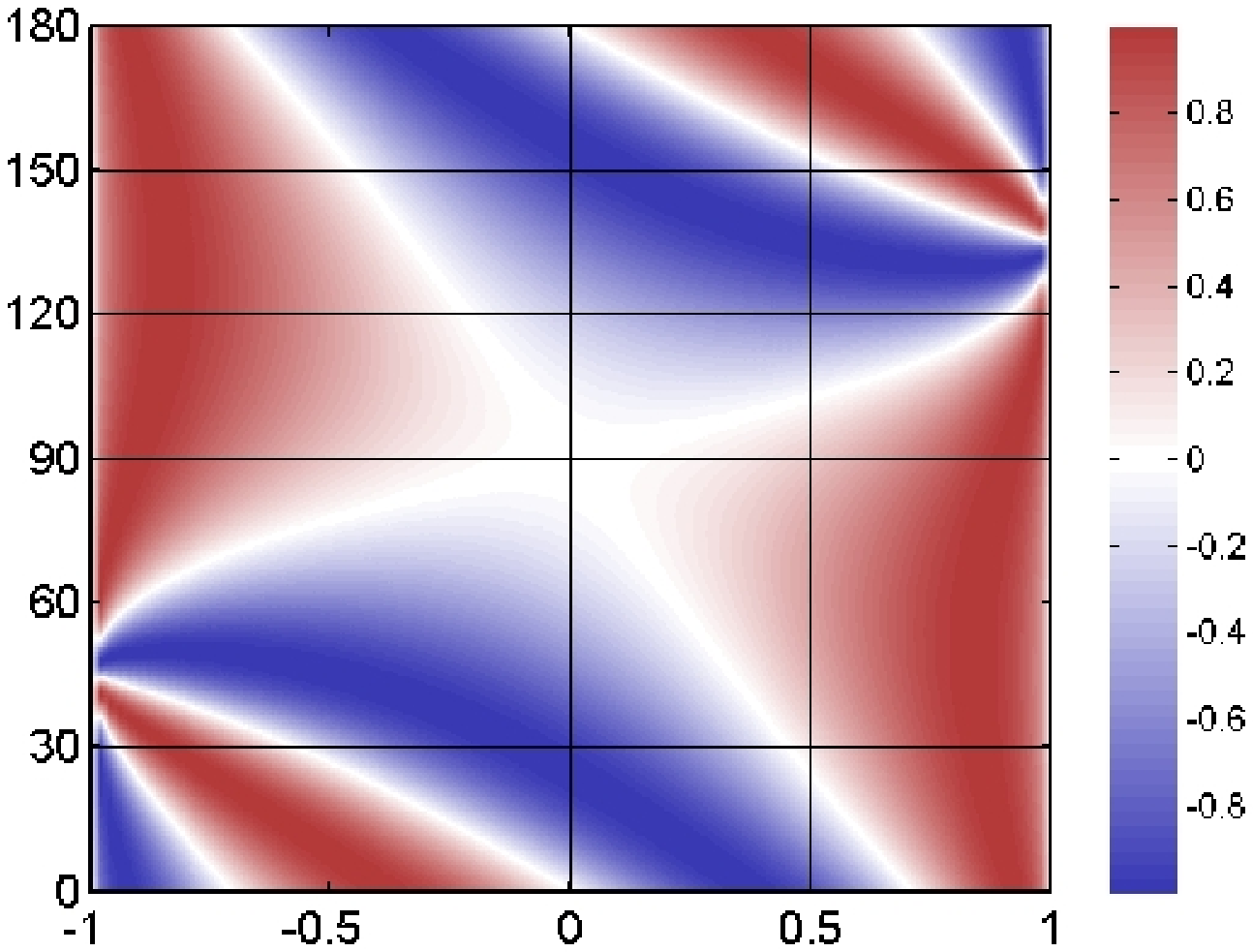}
\includegraphics[width=0.42\textwidth]{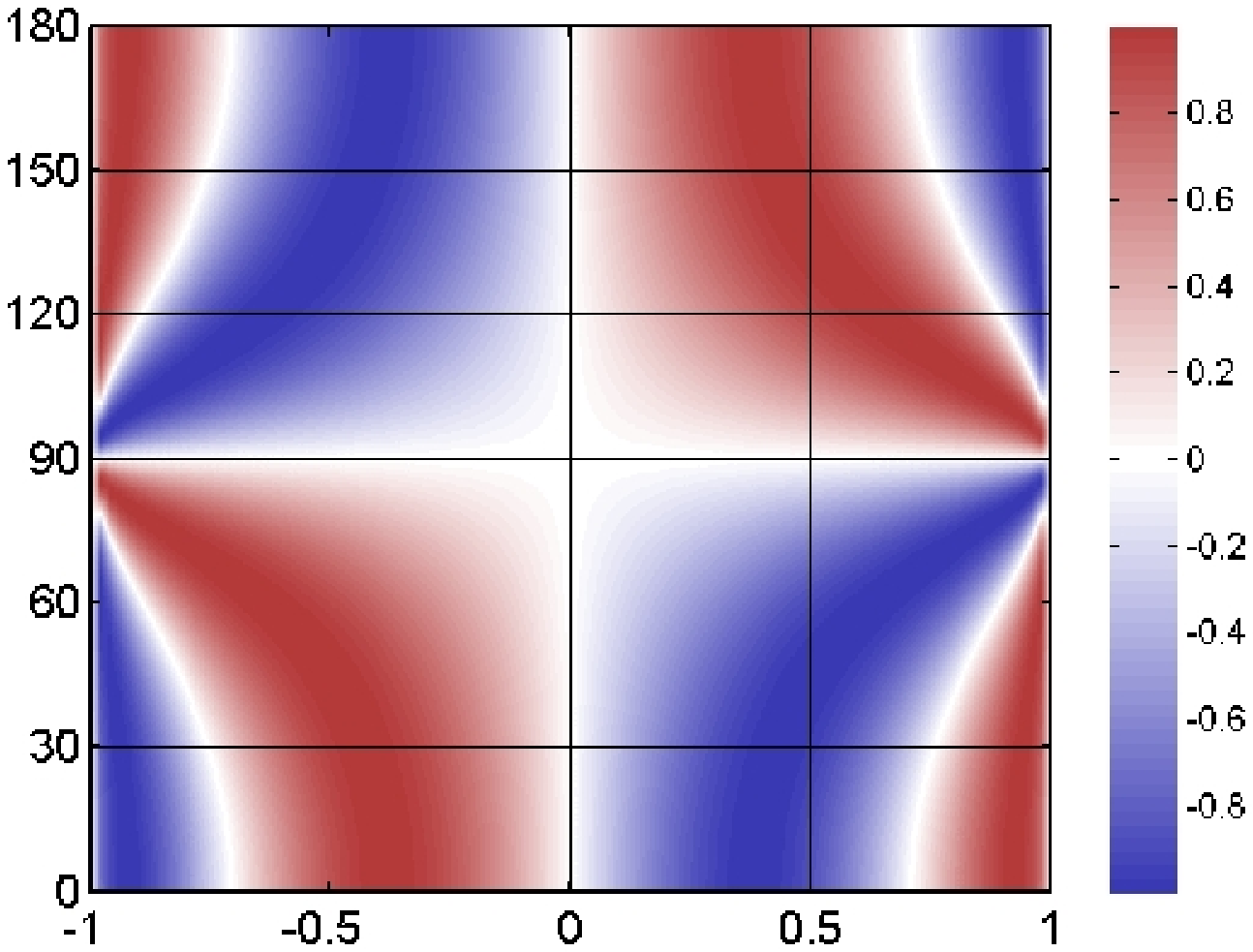}
\label{fig:sign_plots}
\caption{Diagrams showing the sign of CP generated by Faraday
conversion in a helical jet {\bf B} field as a function of the relative
location across the jet (horizontal axis running from $-1$ at left to $+1$
at right, 0 corresponds to the ``spine'' of the jet) and the angle at which 
the jet is viewed in the jet rest frame 
(vertical axis running from $0^{\circ}$ at bottom to $180^{\circ}$ at top; 
$0^{\circ}, 90^{\circ}$ and $180^{\circ}$ correspond to an observer viewing 
the jet ``head-on'', ``side-on'' and ``tail-on'', respectively). The left 
column shows diagrams
for right-handed helices with pitch angles $5^{\circ}$, $20^{\circ}$, 
$40^{\circ}$ and $45^{\circ}$ (top to bottom), while the right
column shows diagrams for right-handed helices with pitch angles of $50^{\circ}$ 
and $70^{\circ}$, a left-handed helix with pitch angle $70^{\circ}$ and
the limiting case of a right-handed helix with pitch angle $90^{\circ}$. Red 
and blue correspond to positive and negative CP.}
\end{figure*}

To address this question in a first approximation, we have constructed
two-dimensional plots of the expected CP sign for a hollow, right-handed
helical {\bf B} field as a function of distance from
the jet axis and viewing angle in the jet rest frame, shown in
Fig.~6. Note that there is some physical basis for adopting such a
model, since the thermal plasma giving rise to the transverse RM 
gradients appears to be external to the bulk of the jet volume in at
least some cases (Sikora et al. 2005). In Fig.~6, the distance 
from the jet axis
is indicated in arbitrary units, with 0 corresonding to the position of the
jet axis (``spine'') and $\pm 1$ corresponding to the two edges of the jet.
Regions of positive CP are coloured red and regions of negative CP blue.
Regions predicted not to give rise to CP because the projected angle
between the background and foreground fields is 0 or a multiple of
$90^{\circ}$ are white, while regions corresponding to the intermediate
``optimal'' angles for generating CP via Faraday conversion, near odd
multiples of $45^{\circ}$, are coloured most intensely. These diagrams are
calculated based purely on the angle between the background field
${\bf B}_{gen}$ and foreground field ${\bf B}_{conv}$ for a helical field
with pitch angle $\psi$ as seen by an observer viewing the jet at an angle
$\theta$ projected onto the sky (this formula corresponds to a right-handed 
helix; in the analogous expression for a left-handed helix, the pitch angle 
$\psi$ is replaced by $-\psi$):

\begin{eqnarray*}
\Phi & = & 2\arctan \left[\frac{\sqrt{1-x^2}\sin\psi}{\sin\theta\cos\psi -
x\cos\delta\sin\psi}\right]\\
\end{eqnarray*}

\noindent
Thus, it is important to bear in mind that they do not represent the
actual intensity of the CP, merely the sign that should be produced for
a specified viewing angle, pitch angle and relative distance
from the jet axis. For comparison, we also show the CP sign distribution 
across the jet as a function of viewing angle for a pitch angle of $70^{\circ}$
and a left-handed helical field. Note that, in this simple treatment, 
these images are point-symmetric.

While these diagrams do not take into account a number of potentially
significant effects, such as effects due to the propagation of the synchrotron
radiation emitted at the ``back'' of the jet along the line of sight through
the jet volume on its way to the observer, they nevertheless provide a useful
overview of the expected robustness of our approach. We plan in future to
carry out more complete radiative-transfer calculations of the expected 
CP distributions across the jet as a function of pitch angle and viewing angle.

The approach we have used will be relatively robust if the predicted CP sign
is constant over a large fraction of the jet cross section and for a fairly
wide range of viewing angles. We can see that this will be the case for 
essentially all pitch angles that are not too close to those values that
yield zero CP. There is an interesting selection effect here that 
helps make our approach relatively robust -- the pitch angles for which
there is a relatively small range of viewing angles corresponding to the 
CP sign observed for a viewing angle of $90^{\circ}$ in the jet frame are
also those that give rise to relatively weak CP. Thus, given the weakness
of the CP signal, the AGN from which we detect parsec-scale CP will tend to
be those with jets whose pitch angles are far from these values, so that 
the inferred CP sign is constant over a wide range of viewing angles.

\subsection{Expected Overall Robustness of the Approach}

Despite the success of our approach for the eight AGN considered here,
there are limitations to its robustness. We argue above that the dependence
of the CP sign on the viewing angle is not likely to play the most important 
role. We note as well that this first simple analysis has not included 
the effects of possible modest Faraday rotation in the jet volume, which
will affect the angle between the polarization {\bf E} vector and conversion
{\bf B} field. However, the weakest link in the procedure we have outlined 
is probably 
inferring the pitch angle regime for the helical jet {\bf B} field based
on the observed polarization structure. In particular, there are other physical
mechanisms that could give rise to a predominance of orthogonal or
longitudinal {\bf B} fields in AGN jets. Orthogonal fields can be generated
by shock compression, and the longitudinal {\bf B}-field component may be
enhanced during shear interactions between the jet and the medium through
which it flows. If one or both of these mechanisms operate, the associated
{\bf B}-field structure will be superposed on the ``intrinsic'' {\bf B}-field 
structure associated with the helical field, possibly leading to incorrect
conclusions about the pitch-angle regime for the helical field. However, the 
fact that our approach also requires that there be a transverse RM gradient
across the jet, thereby providing direct evidence for a predominant helical 
{\bf B} field, may help reduce ambiguity in the inferred pitch-angle regime
due to these effects.

\subsection{Implications of Results for the Jet ``Seed'' Field}

As is shown in 
Fig.~2, a right-handed helix with a South poloidal 
component will display the same sense of RM gradient as a left-handed helix 
with a North poloidal component. This means, for example, that the 
helicity we would infer in this case based on the observed transverse RM 
gradient would be incorrect if we assumed the jet base represented a 
\emph{North} magnetic pole, but in fact it represented a \emph{South} pole.

Therefore, if statistically equal numbers of 
the approaching jets in the 8 AGN considered here corresponded 
to North and South magnetic poles, the sense of the poloidal fields that
yielded agreement between the predicted and observed CP signs would have
been North in roughly half the cases and South in roughly half the cases. 
Since we have
no other way to verify the sense of the poloidal field observationally,
it would not have been possible to draw conclusions about whether, in fact,
conversion in a helical-field geometry was the origin of the observed CP. 
However, our expected CP signs for South poloidal fields matched the 
observed values in 8 out of 8 cases; the probability of this happening
just by chance is less than 1\% (it is equivalent to the probability of
flipping a coin 8 times and getting ``heads'' all 8 times).

This leads us to conclude that our results provide evidence for a predominance
of South poloidal components among the jet {\bf B} fields. 
This does not seem plausible if the seed fields that are transformed into a 
helical-like field by the jet outflow and rotation of the central objects are 
essentially dipolar.  In this case, each black hole would have one jet with a 
North poloidal field and one with a South poloidal field, which should 
be randomly oriented in space, so that there are equal probabilities that the 
approaching jet corresponds to the South and North poles of the black hole.
However, another possibility is that the seed field generated
by the rotating central black hole is actually quadrupolar-like, with two
North poles or two South poles, such that the magnetic-field lines close
through the equatorial plane. In fact, it has recently been suggested 
that such a quadrupolar {\bf B}-field may more likely be realized
for black-holes/accretion-disc systems in AGN  than a dipolar field
(Blandford 2007). Our results would be consistent 
with this type of initial {\bf B}-field configuration if there were some
reason for a predominance of quadrupolar initial {\bf B}-field configurations
with two South poles, rather than two North poles. At present, however,
we are not aware of any possible physical origin for such an asymmetry.

\section{Conclusion}

Despite extensive analyses of the degree of CP in previous studies, we are
the first to consider the {\em sign} of the parsec-scale CP of AGN.
The new transverse Faraday-rotation gradient and parsec-scale CP detections
reported here have enabled us to compile a
list of 8 AGN with both CP detections and
direct evidence for helical jet {\bf B} fields.
The perfect agreement between the observed CP signs and the CP signs expected if 
the CP is generated by Faraday conversion in helical jet fields whose 
poloidal component is oriented opposite to the jet
conclusively demonstrates that the observed CP is intrinsically
related to the helical jet {\bf B} fields in these objects. Although the
possibility of conversion-generated CP has been considered in a number of
theoretical studies, this represents the first direct 
\emph{observational} evidence pointing toward
this specific CP-generation mechanism in the more than three decades since the
earliest studies of CP in AGN.

This result has far-reaching implications for our understanding of
the processes occurring in AGN jets, since it suggests that helical
jet {\bf B} fields are common in AGN, possibly even ubiquitous. Indeed,
such helical {\bf B} fields can be produced very naturally through
the ``winding up'' of a seed field by rotation of the central black
hole and accretion disc, combined with the jet outflow (e.g. Meier
et al. 2001; Kato et al. 2004; McKinney 2006). The generation of a helical
field requires the presence of a significant poloidal field, which
can come about due to the jet outflow. The presence
of helical jet {\bf B} fields provides a means to collimate the jets,
and demands that the jets carry current (e.g. Lovelace et al. 2002,
Tsinganos \& Bogovalov 2002). Thus, it is clear that we
must view AGN jets as fundamentally electromagnetic structures, although
it is less clear from the available observations whether or not the jets
are actually dynamically dominated by the associated electromagnetic
forces.

Our results also appear to provide evidence that the initial ``seed''
fields of the central black holes are typically quadrupolar rather than 
dipolar, with a predominance of quadrupolar fields with two South poles.
Numerical simulations of the joint action of rotation and outflow on 
an initial quadrupolar field would be helpful in determining if there 
is a theoretical basis for this possible asymmetry between these two 
field configurations. 
An interesting observational distinction between dipolar and quadrupolar
field configurations for the central black holes was recently pointed out
by Blandford (2007): in the case of a helical jet {\bf B} field generated
from a dipolar seed field, the transverse Faraday-rotation gradients for the
approaching jet and receding counterjet should have \emph{opposite} directions,
while these two gradients will be in the \emph{same} direction if the seed
field that is ``wound up'' by the rotation of the system is quadrupolar. 
The observational challenge is therefore to identify AGN in which 
both total intensity and polarization are detected in both the jet and 
counterjet  on scales on which the transverse Faraday-rotation gradients
due to the helical jet {\bf B} fields are reliably detected, and to carry
out sensitive Faraday-rotation measurements for such objects (the apparent
absence of clear RM gradients in FRI jets on kiloparsec scales (e.g. Laing 
\& Bridle 2007) may indicate that the systematic transverse RM gradients 
are disrupted by chaotic or turbulent motions on these scales).

We plan to analyze multi-wavelength VLBA polarization observations 
for several additional AGN with detected parsec-scale CP in order to
search for transverse Faraday-rotation gradients in these objects,
with the aim of further investigating the
relationship between the linear-polarization structures, {\bf B}-field
helicities and observed CP in AGN.  We also plan to consider
the possible influence of modest amounts of Faraday rotation on the
observed CP in a future paper.

\section{Acknowledgements}

We thank T. V. Cawthorne, D. Meier, N. O'Murchadha and J. F. C. Wardle
for useful and interesting discussions of this work. We are also grateful 
to the referee, Thomas Beckert, for pertinent and clearly explained comments 
that have appreciably improved this paper.






\end{document}